# Digital quantum simulation of NMR experiments

Kushal Seetharam*,[1,2] Debopriyo Biswas,[3,4] Crystal Noel,[3,4] Andrew Risinger,[4] Daiwei Zhu,[4] Or Katz,[3] Sambuddha Chattopadhyay,[2] Marko Cetina,[4,5] Christopher Monroe,[3,4,6] Eugene Demler,[7] and Dries Sels[8,9]

[1]*Department of Electrical Engineering, Massachusetts Institute of Technology, Cambridge, Massachusetts 02139, USA*
[2]*Department of Physics, Harvard University, Cambridge, MA, 02138, USA*
[3]*Department of Electrical and Computer Engineering, Department of Physics, Duke Quantum Center, Duke University, Durham, North Carolina 27708, USA*
[4]*Joint Quantum Institute, Department of Physics, University of Maryland, College Park, Maryland 20742, USA*
[5]*Department of Physics, Duke Quantum Center, Duke University, Durham, North Carolina 27708, USA*
[6]*IonQ, Inc., College Park, Maryland 20740, USA*
[7]*Institute for Theoretical Physics, ETH Zürich, 8093 Zürich, Switzerland*
[8]*Department of Physics, New York University, New York, NY, 10003, USA*
[9]*Center for Computational Quantum Physics, Flatiron Institute, New York, NY, 10010, USA*

Simulations of nuclear magnetic resonance (NMR) experiments can be an important tool for extracting information about molecular structure and optimizing experimental protocols, but are often intractable on classical computers for large molecules such as proteins, and for protocols such as zero-field NMR. We demonstrate the first quantum simulation of a NMR spectrum, computing the zero-field spectrum of the methyl group of acetonitrile using four qubits of a trapped-ion quantum computer. We reduce the sampling cost of the quantum simulation by an order of magnitude using compressed sensing techniques. We show how the intrinsic decoherence of NMR systems may enable the zero-field simulation of classically hard molecules on relatively near-term quantum hardware, and discuss how the experimentally-demonstrated quantum algorithm can be used to efficiently simulate scientifically and technologically relevant solid-state NMR experiments on more mature devices. Our work opens a practical application for quantum computation.

## INTRODUCTION

Nuclear magnetic resonance (NMR) spectroscopy is a widely used tool in structural biology and materials chemistry, providing insight into the structure, dynamics, reaction state, and chemical environment of both liquid-state and solid-state systems [1]. Systems studied with liquid-state NMR range from to small biomolecules such as metabolites to large biomacromolecules such as proteins, nucleic acids, carbohydrates, and lipids, while solid-state NMR can help characterize various rigid and semi-rigid systems such as membrane proteins, amyloid fibrils, polymers, battery materials, photovoltaic perovskites, solid state catalysts, and metal-organic frameworks [2, 3]. Despite their versatility, NMR experiments can be difficult to interpret for large liquid- and solid-state systems. Current approaches typically combine sophisticated experimental protocols that simplify the nuclear spin dynamics at the heart of the experiment with heuristic formulas to infer the information of interest. When available, numerical simulation of the spin dynamics can provide an important tool in the analysis and optimization of conventional high-field liquid-state experiments on large biomolecules [4], emerging zero-field experiments on small biomolecules [5, 6], and many high-field solid-state experiments [2]. For example, simulation can be used to optimize experimental protocols and pulse sequences, validate chemical and structural information extracted from heuristic formulas applied to the experiment, and open the door to novel structure determination paradigms [4].

Numerical simulations of NMR experiments, however, can be very challenging to perform on classical computers when either the dynamics itself is hard, meaning correlations generated by the quantum dynamics of the system becomes difficult to keep track of (see the Supplementary Materials), or when an ensemble ('powder') average over system orientations is performed [2]. Liquid-state NMR exhibits complex dynamics in zero-field experiments dominated by coherent exchange interactions even for relatively small biomolecules, and in high-field experiments dominated by dipolar-relaxation for large biomacromolecules. Solid-state NMR, dominated by coherent dipolar interactions, often exhibits hard dynamics and necessitates powder averaging over differently oriented grains in the sample.

Quantum computers and simulators, themselves described as an effective system of spins, are naturally suited to simulate the dynamics of spin systems. Indeed, such simulations may be the first practical application of quantum computers to achieve a speed-up compared to classical computers [7]. Quantum hardware may therefore enable the simulation of NMR experiments with complex dynamics [8]. These spin dynamics simulations can enable inference of Hamiltonian parameters encoding the chemical and structural information of interest in NMR experiments. Liquid-state zero-field NMR experiments on relatively small molecules may be the earliest achievable context where quantum computers can display an advantage over classical computers. Solid-state

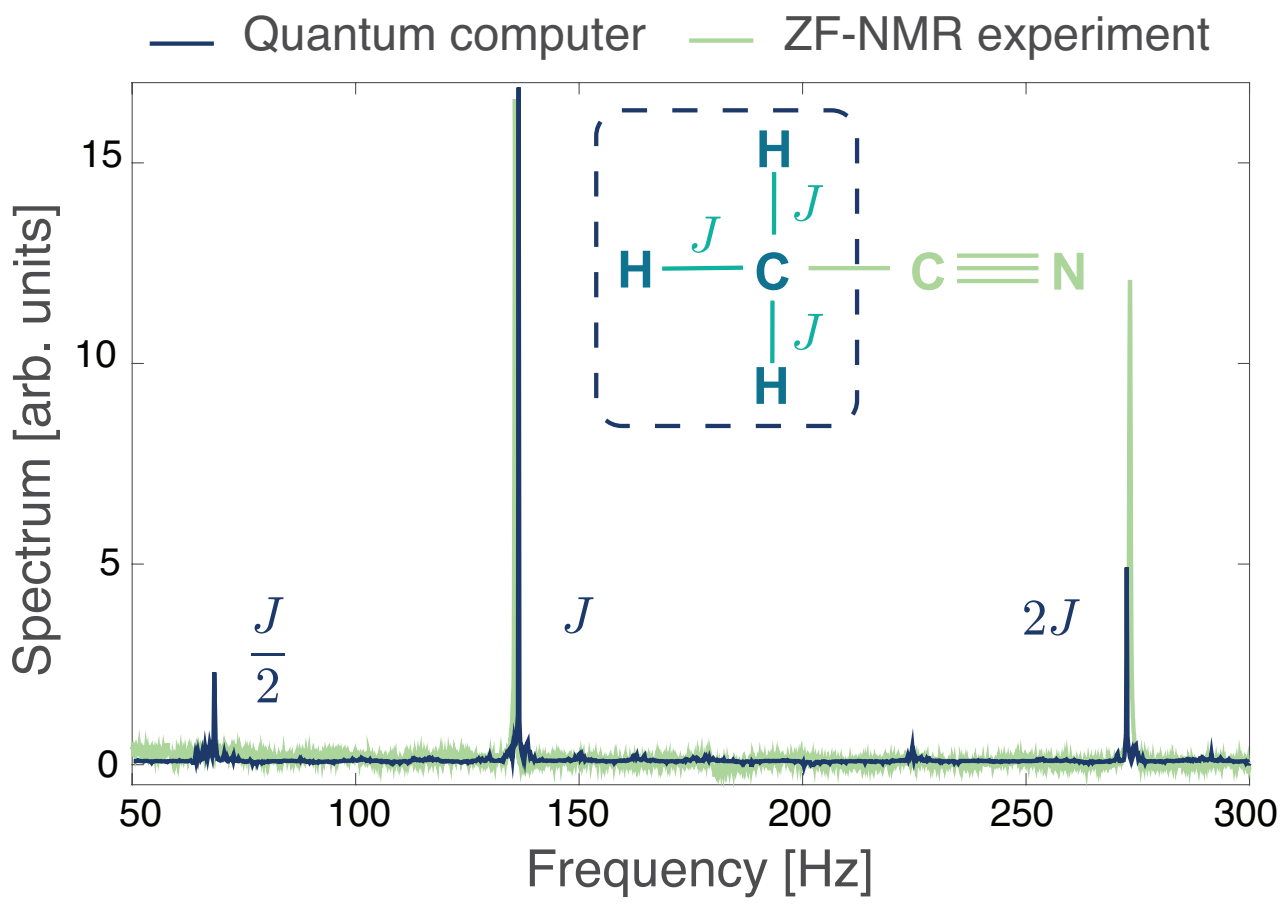

Figure 1: **Liquid-state NMR spectrum computed on quantum hardware.** Zero-field spectrum of acetonitrile computed on an ion-trap quantum computer (blue curve) compared with the NMR experiment (green curve) performed in Ref. [10]. The inset shows the chemical structure of acetonitrile, highlighting the methyl group that was probed in the experiment.

NMR, on the other hand, is a more widely relevant context where quantum computers may have greater practical utility, but would require substantial improvements to current quantum hardware to simulate the large systems typically investigated in experiments.

Here, we simulate a zero-field NMR experiment on a trapped-ion quantum computer [9]. The quantum computer implements a carefully compiled sequence of unitary rotations and entangling interactions on $^{171}Yb^+$ ion qubits, to implement a digital quantum circuit that emulates the NMR experiment (see the Supplementary Materials). Specifically, we compute the spectrum of selectively isotope labeled acetonitrile, with four NMR-active nuclear spins, and show that the resonance frequencies in the spectrum quantitatively match the experimental NMR data from Ref. [10], while the peak intensities match in their ordering. We obtain high spectral resolution within the resource limitations of the trapped-ion device by exploiting compressed sensing techniques [11] and a state-of-the-art quantum circuit synthesis algorithm [12]. These techniques, which leverage the power of the individual qubit control of a digital quantum computer, enable more versatile simulations of NMR spectra than would be possible on typical analog quantum simulators which seek to natively implement the desired dynamics [13]. They reduce the resource cost for simulating classically hard NMR systems substantially, and are likely to prove useful in quantum simulations of hard systems that appear in quantum chemistry and condensed matter physics [14]. We give resource estimates for quantum simulations of relatively small molecules whose zero-field NMR spectra are practically challenging to simulate on classical computers, showing how the dephasing commonly present in nuclear spin dynamics may enable such simulations on relatively near-term quantum devices. We then discuss how more mature quantum hardware may enable efficient analysis of large systems such as membrane proteins and battery materials studied using solid-state NMR experiments.

## RESULTS

A NMR experiment involves polarizing the nuclear spins of a sample via an external magnetic field or a chemical process, letting the spins evolve in time, and measuring the average magnetization of the system. The measured time-dependent magnetization is called the free induction decay (FID), and its Fourier transform yields the NMR spectrum. Letting the operators $\{\mathbf{S}_i\}$ represent the nuclear spins, the initial state of the system when polarized via either a large magnetic field or a chemical process can be described as $\rho_0 \approx I + \lambda \tilde{S}^z_{\text{tot}}$, where $I$ is the identity operator and $\tilde{S}^z_{\text{tot}} = \sum_i \gamma_i S^z_i$, with $\gamma_i$ being the gyromagnetic ratio of the nuclear isotope $i$. In the case of a 1D NMR experiment, the measured FID corresponds to the quantity

$$\text{FID}\,(t) = \text{Tr}\left[U\,(t)^\dagger\, \tilde{S}^z_{\text{tot}} U\,(t)\, \tilde{S}^z_{\text{tot}}\right], \tag{1}$$

where $U\,(t) = \exp\,(-iHt/\hbar)$ produces the time-evolution of the system generated by a Hamiltonian $H$. Most NMR experiments can be modeled by Eq. (1) with different Hamiltonians and, in the case of multi-dimensional protocols, global spin rotations that modify the time-evolution [1]. The evolution of liquid-state molecular samples is typically well captured by

$$H = \sum_{i,j} 2\pi J_{ij} \mathbf{S}_i \cdot \mathbf{S}_j + \sum_i \omega_i S^x_i, \tag{2}$$

where we have taken Planck's constant $\hbar = 1$. The $J$-couplings $\{J_{ij}\}$ characterize the strength of bond-mediated exchange interactions and the chemical shifts $\{\omega_i\}$ represent local magnetic screening around nuclei in different chemical environments in response to an external magnetic field [1].

Zero-field NMR protocols avoid the external field, opening the possibility of portable and cheaper experiments as they obviate the need for cryogenically cooled superconducting magnets [5, 6, 10]. Without a large background field, however, the interactions between spins become dominant. Therefore, a notable limitation of zero-field protocols is that their spectra are hard to interpret without access to simulations of the NMR experiment, which can be rendered classically intractable for even relatively small molecules [15]. High-field liquid-state protocols, in contrast, are typically easy to simulate as the scalar $\mathbf{S}_i \cdot \mathbf{S}_j$ interaction in Eq. (2) reduces to its secular component $S^z_i S^z_j$, which substantially simplifies the complexity of the dynamics.

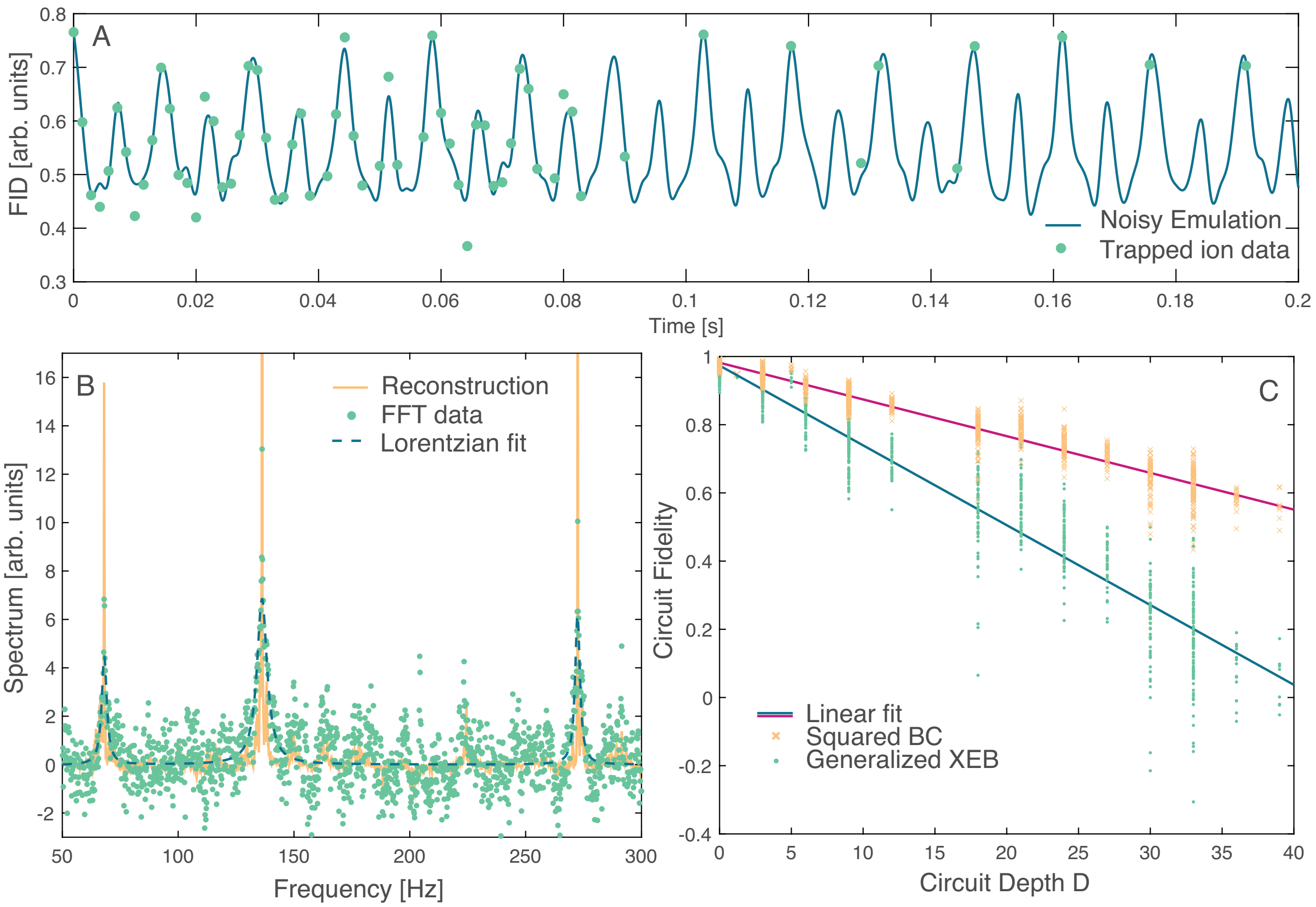

Figure 2: **Compressed sensing reconstruction** & **Benchmarking.** **A** Comparison of the FID of a noisy quantum circuit emulation (blue line) and the non-uniform, sparsely sampled points experimentally measured on the trapped-ion quantum computer (green circles). The noise is modeled by two-qubit gates subject to both amplitude and phase damping with rates 0.005s and 0.035s respectively. **B** NMR spectrum extracted from the digital quantum simulation, where the spectrum is the real part of the Fourier transform of the FID. Green dots show the spectrum after replacing unsampled points of the FID with zeros. Dashed blue line shows the best (under $\ell_1$-norm) Lorentzian fits to this zero-padded data. Solid yellow line shows the reconstructed spectrum after applying the IST-S algorithm. The y-axis is rescaled (zoomed-in) compared to Fig. 1 to make the features more visible. **C** Fidelity of quantum simulation. The yellow crosses show the squared Bhattacharyya coefficient and the green dots show a generalized cross entropy benchmark as a function of the circuit depth measured in the number of two-qubit gates.

We compute the zero-field spectrum of acetonitrile, a compound which is commonly used as an industrial solvent. The molecule is isotope labeled to have four NMR-active spin-1/2 nuclei, a $^{13}$C and three $^{1}$H, that make up a methyl group (see inset in Fig. 1). There are three non-zero $J$-couplings, corresponding to the three $^{13}$C$-^{1}$H bonds, all with value $J = 136.2$ Hz. The FID signal of Eq. (1) can be computed on a quantum computer using four qubits by initializing the system qubits in basis states with a positive average magnetization, enacting time-evolution under the Hamiltonian via an appropriate quantum circuit, Eq. (2), and then measuring the average magnetization of the system. We write this measurable as

$$\mathrm{FID}\,(t) = \sum_{\tilde{m}_n > 0} \tilde{m}_n \, \langle \tilde{m}_n(t) | \tilde{S}^z_{\mathrm{tot}} | \tilde{m}_n(t) \rangle \,, \qquad (3)$$

where $\{|\tilde{m}_n\rangle\,;\tilde{m}_n\}$ are the eigenstates and eigenvalues of $\tilde{S}^z_{\mathrm{tot}}$, and $|\tilde{m}_n(t)\rangle = U(t)\,|\tilde{m}_n\rangle$. For a system of $N$ spins, the sum in Eq. (3) can have a number of terms that scales exponentially with $N$, naively negating the quantum computational advantage. However, the variance of the estimator is much smaller than exponential and can be bounded by $N^2$ due to the bounded operator norm of the total magnetization being measured; at most $N^2$ terms of the sum can thus be used to estimate the FID via either uniform or importance sampling [8]. The described quantum algorithm can be straightforwardly generalized to simulate a diverse array of NMR experiments, including protocols with different initial conditions, multiple dimensions, solid-state samples, and different isotope labeling. Multidimensional protocols can be simulated by inserting single-qubit rotations into the time-evolution of $U(t)$, and solid-state experiments can be simulated by including a dipolar interaction term in the Hamiltonian of Eq. (2) (see the Supplementary Materials). Isotope labeling is incorporated by the choice of basis states that the qubits are prepared in and by the choice of qubits to measure at the end of the simulation.

Figure 1 shows the spectrum we compute on an ion trap quantum computer in comparison with the seminal zero-field NMR experiment of Ref. [10]. We see that the quantum computation accurately reproduces the resonances at frequencies $J$ and $2J$. Specifically, the corresponding resonance frequencies extracted from the quantum simulation are $136.20{\pm}0.09$ Hz and $272.41{\pm}0.09$ Hz, which are within $1\sigma$ of the exact frequencies of 136.2 Hz and 272.4 Hz. The extracted resonance frequency uncertainty is Fourier limited by the total acquisition time; a Lorentzian fit to the reconstructed peaks results in a width smaller than the frequency grid spacing. We therefore take half the grid spacing as the uncertainty. Given that the zero-field NMR experiment can only resolve the spectral peaks within 0.1 Hz [10], we demonstrate that quantum computers can accurately simulate NMR data with a quality comparable to real experiments.

The spectrum computed on the quantum computer has peak intensities that match the ordering of peaks in Ref. [10], but has a quantitative mismatch with spectral weight transferred from the resonance at $2J$ to an additional resonance at $J/2$ that is not present in the NMR experiment. This additional spectral peak arises from a combination of errors in the quantum computer and the high-symmetry of the molecule, which induces dynamical recurrences that are captured by the specific method we use to synthesize the time-evolution circuits. Such artifacts are unlikely to appear in classically intractable NMR simulations whose large, strongly correlated molecules typically do not exhibit many-body revivals. Furthermore, we provide a simple method to remove artifact peaks in future experiments even for the small, highly symmetric systems where they may occur (see the Supplementary Materials).

In order to calculate the spectrum, we first compute the FID, Eq. (3), at a non-uniform random sampling of time points lower than the Nyquist rate. We synthesize the time-evolution quantum circuits using the numerical optimization algorithm in Ref. [12] after tailoring it to the gate set and qubit topology of the trapped ion device (see the Supplementary Materials). This numerical synthesis procedure efficiently produces low-depth circuits but is limited to a small number of qubits because it searches of the full space of possibly unitaries. It can, however, be a useful tool when simulating larger systems by combining it with a cluster Trotterization method, as described in the Supplementary material.

The undersampled FID measured in experiment is reconstructed into a spectrum by a recovery algorithm which assumes that the time domain signal is sparse in the frequency domain. These two steps – non-uniform sampling (NUS) and spectral reconstruction – form the basis of compressed sensing. Compressed sensing techniques have their root in information theory [16], but have been further developed in the experimental NMR community where they can drastically reduce the data collection burden [11]. While these techniques have recently been used in quantum sensing [17], we demonstrate their use in quantum simulation experiments to similarly reduce the computational cost [14]. In Fig. 2A we plot a noisy emulation of the ion trap experiment at all values of the uniform dense time grid and compare to the NUS points that were actually collected in the experiment. Noise is modeled with amplitude and phase damping channels acting on each two-qubit gate, with the rate of each channel determined by fitting the emulation to experimental data. Experimental data was collected up to times $t = 6s$ (see the Supplementary Materials), but is only shown up to $t = 0.2s$ to allow a clear comparison to the noisy emulation. We use a sine-weighted Poisson gap NUS schedule that is dense at short times as it has been shown to reduce reconstruction artifacts [18]. Figure 2B shows the spectrum resulting from Fourier transforming the experimental data before running the reconstruction algorithm. We see that the signal-to-noise ratio in this raw spectrum is poor due to NUS artifacts, with a Lorentzian fit to the peaks resulting in an uncertainty of approximately 1 Hz. The same spectrum is shown after we run the iterative soft thresholding (IST-S) reconstruction algorithm; the signal-to-noise is dramatically improved, with the uncertainty reducing by an order of magnitude to approximately 0.1 Hz. The reconstructed spectrum matches the spectrum resulting from fully sampled noisy emulation (see the Supplementary Materials). Experimentally, only 102 out of the 4096 time points were collected, indicating that compressed sensing reduced the computational burden of the experiment by more than a factor of 40. This reduction is particularly crucial for experiments with slow repetition rates. We note that compressed sensing techniques will remain applicable to quantum simulation of NMR experiments on large, classically-intractable systems well-beyond the small molecule demonstrated in this work. The multidimensional NMR protocols used to study such systems are designed to spread spectral weight across different dimensions in order to improve interpretability of the spectrum, thus manifesting the frequency-domain sparsity required for compressed sensing.

In Fig. 2C, we asses the quality of the trapped-ion simulation by comparing the outputs of all 102 circuits ($\times$ 8 initial states) with the ideal outputs resulting from a noiseless circuit emulation. These synthesized circuits, each corresponding to a time $t$, have varying circuit depths according to the entanglement generated in the system at that time (see the Supplementary Materials). The Bhattacharyya coefficient (BC), which provides an upper bound for the fidelity of the prepared quantum state (see the Supplementary Materials), indicates that a typical two-qubit gate operation was enacted with fidelity at most 98.9%. The BC is an informative metric for states with high fidelities, but it saturates to a value of 0.5 for the random states that the system tends to after decoherence runs its course. We therefore also examine a generalized cross entropy benchmark (gXEB) introduced in Ref. [19], which is a better estimate for the fidelity. The gXEB yields an estimate of 97.7% fidelity per operation enacted in the trapped-ion experiment. It should be noted that, similar to the XEB [20], the gXEB can be negative.

While the present experiment is performed on state-of-the-art quantum hardware, it is still easily tractable on a classical computer. In order to elucidate the hardware resources required to scale quantum simulations to classically hard zero-field NMR experiments, we examine three challenging systems that are at the border of what is classically simulable [21, 22]. The compounds are depicted in Fig. 3A and are taken from the example set of Spinach [23], an advanced classical simulation package that leverages decoherence in the NMR experiment to make the compuation more efficient [4]. Each system can be simulated on a classical computer in several hours, provided access to 32 CPU cores, 128 GB RAM, and a graphics card as powerful as the Titan V. The interaction graphs characterizing the molecules' nuclear spin Hamiltonians have a compact structure, and are composed of strongly interacting clusters of four to seven spins which are weakly connected to other clusters. The compact nature of the interaction graphs—which give rise to rapidly spreading strong correlations—makes these systems hard to simulate on a classical computer, even though these NMR experiments can be described without the long-range dipolar interactions that are central to other challenging NMR experiments such as solid-state NMR.

We estimate the resources required to simulate these systems using a quantum computer by using product formulas to prescribe circuits that implement time-evolution under the Hamiltonian of Eq. (2). While there are many quantum algorithms that implement quantum dynamics, product formulas are considered to have the lowest resource overhead and be most suitable for early quantum devices [7, 24]. We exploit both the cluster structure of the nuclear interactions as well as inherent dephasing in the NMR experiment to further reduce the cost (see the Supplementary Materials). With the total number of qubits fixed to the number of nuclear spins, the performance crucially depends on the accuracy of the Hamiltonian engineering (Trotter error) as well as the fidelity with which each operation can be performed.

In Fig. 3B, we plot the achievable linewidth $\Delta f$ of the NMR spectrum as a function of the circuit depth $D$ for quantum computers with various levels of noise, here we model the decoherence and dephasing with a single parameter $F$ which encapsulates the average fidelity of a gate. We assume the time-evolution quantum circuits are designed using a clustered first order product formula (see the Supplementary Materials). We define the circuit depth as the number of (arbitrarily connected) two-qubit gates, as available in ion trap quantum computers [9]. We observe a $1/\sqrt{D}$ scaling, reminiscent of the standard quantum limit, up to a critical depth where the decoherence of the quantum computer takes over. The improvement in resolution is due to a decrease in the Trotter error with circuit depth. At any given value of the gate fidelity $F$ there is an optimal circuit depth $\sim 1/\log(1/F)$ arising from a competition between algorithmic error and decoherence, resulting in linewidth $\Delta f \sim \sqrt{\log(1/F)}$. Fig. 3C depicts the expected optimal linewidth for all the molecules considered in this work. While we clearly observe that the larger molecules from Fig. 3A are considerably harder to simulate than the four spin methyl group that was computed here, it should be noted that these curves are expected to saturate for Hamiltonians corresponding to clustered molecules. To simulate the phosphorus cluster (Fig. 3A(iii)) to the same level as the physical NMR experiment, we expect to require circuits of $O(10^5)$ gates with a typical gate infidelity of $O(10^{-4})$, an infidelity that is two orders of magnitude better than the present experiment. Such infidelities have been achieved in small trapped-ion systems [25, 26], and future scaling strategies hold great promise for reaching the above performance metrics [27].

## DISCUSSION

Zero-field NMR experiments on molecules of similar size to those in Fig. 3A may thus be a context where prospective near-term quantum devices can show an advantage over classical computers. A more practically compelling context, however, is solid-state NMR. The same quantum algorithm described by Eq. 1 can be used to efficiently simulate the dynamics of solid-state NMR experiments after adding a dipolar interaction term to the Hamiltonian in Eq. 2 (see the Supplementary Materials). Magic angle spinning protocols can be modeled by endowing this dipolar term with time-dependant coefficients. Simulating solid-state NMR experiments, however, necessitates performing a powder average over $10^3$ to $10^4$ orientations of the system, each corresponding to an independent FID that must be computed. On classical computers, the simulation of each fixed orientation FID can be challenging for large systems such as membrane proteins or battery materials, thus making the overhead of powder averaged calculations onerous even after exploiting parallelization. The resource cost for quantum simulation of the powder-averaged FID for these systems, however, is roughly the same as the cost for computing a single FID corresponding to a fixed orientation.

Specifically, the sample complexity of the two cases is approximately equivalent on a quantum device. The FID computed for a fixed system orientation, Eq. (1), can be viewed as an estimator for a random variable corresponding to the total magnetization of the system [8]. The powder-averaged FID can be computed by sampling a different orientation every time a term in Eq. (3) is sampled. Letting $\mathbb{E}_{t;\Omega}[M]$ be the expectation value cor-responding to the FID at time $t$ for a system in orientation $\Omega$, the powder-averaged FID is simply $\mathbb{E}_\Omega[\mathbb{E}_{t;\Omega}[M]]$, with the outer expectation corresponding to a classical average over a uniform distribution of orientations. The variance of the estimator for the powder-averaged FID is

$$\mathrm{Var}\left[M\right] = \mathbb{E}_\Omega\left[\mathrm{Var}_{t,\Omega}\left[M\right]\right] + \mathrm{Var}_\Omega\left[\mathbb{E}_{t;\Omega}\left[M\right]\right]. \quad (4)$$

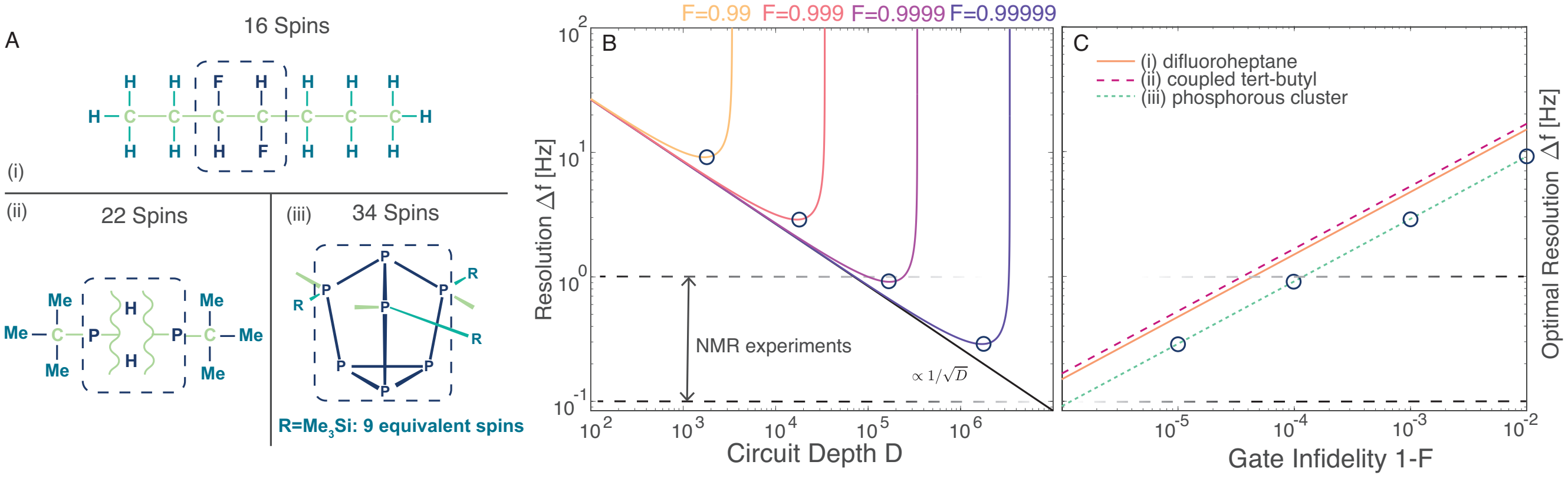

Figure 3: **Scaling up to classically hard liquid-state zero-field NMR simulations.** **A** Chemical structures of (i) anti-3,4-difluoroheptane [21] (ii) a system with two coupled tert-butyl groups and (iii) the $B[ACR_9]_3$ phosphorous system [22]. Light green atoms do not contribute to the NMR signal and dashed boxes indicate strongly interacting clusters who's circuit synthesis can substantially speed up the quantum computation (see the Supplementary Materials). **B** Experimental design curves for $(Me_3Si)_3P_7$ (panel A(iii)), showing $1/\sqrt{D}$ scaling, where $D$ is the circuit depth, of the frequency resolution up to a minimally achievable width set by the decoherence of the quantum computer. The circuit depth is measured by the number of (arbitrarily-connected) two-qubit gates. **C** Optimal resolution for all three molecules. The circles indicate the resolution at optimal circuit depth and the dashed black horizontal lines indicate the resolution accessible in NMR experiments.

The first term is the average quantum shot noise associated with the simulation of a fixed orientation FID, and scales as $N^2$ [8]. The second term captures the classical noise associated with sampling the uniform distribution of orientations comprising the powder average. Empirically, at most $10^3$ - $10^4$ such samples are typically required. For large systems consisting of $N = 10^2$ - $10^4$ spins where quantum devices may prove advantageous over classical computers, the quantum shot noise dominates the classical noise. Therefore, roughly $N^2/\epsilon$ repetitions (shots) of the quantum simulation suffice to achieve a precision $\epsilon$ for both fixed orientation and powder-averaged computations.

The resources required to simulate solid-state NMR experiments on quantum hardware is thus primarily determined by that required to simulate the dynamics of a single orientation of the system. While the hundreds to thousands of simulation-relevant spins in a large solid-state NMR samples may preclude computation of its dynamics on near-term quantum computers without any error correction, more mature quantum hardware may be able to perform the task.

Our zero-field demonstration provides the first proof of principle that quantum computers can simulate NMR spectra within experimental resolution, and the experimentally-demonstrated algorithm may eventually facilitate analysis of solid-state NMR experiments performed on systems of scientific and technological relevance. While scaling quantum NMR simulations to classically intractable systems will be challenging, it should be noted that the resource projections in Fig. 3 are substantially less demanding than most other near-term quantum computing applications [7, 28, 29]. The physical reason behind the reduced resource cost is that dephasing is inherent in the dynamics of nuclear spin systems, with a rate given by the finite line-width of spectral peaks in NMR experiments. Quantum simulations can tolerate decoherence in the quantum device as long as it is less than the dephasing rate of the spin system [8]. We note that the noise characteristics in a quantum simulation platform will not typically be the same as an NMR experiment, as evidenced by the spurious peak seen in Fig. 1. If the noise of the simulation platform is well-understood, it can either be mitigated [30] or made to mimic the noise of the NMR experiment [31, 32]. If the noise of the platform is unknown, techniques such as twirling can be used to turn it into an effective depolarizing channel [33, 34], which is the channel used for the estimates in Fig. 3. In either case, as long as the finite magnitude of noise is sufficiently small, the platform can be used to faithfully simulate NMR experiments.

The hope for an advantage from such digital quantum simulations of NMR lies in time-evolution on quantum hardware scaling with a lower order polynomial of system size than state-of-the-art classical simulation packages such as Spinach. These classical methods also exploit decoherence in the NMR experiment to reduce the state space that is simulated from one scaling exponentially with system size to one scaling polynomially [4].

The subsequent computation, extraction of the NMR spectrum from density matrix time-evolution performed to machine precision using Taylor series expansions, is formally classically tractable, but can prove practically challenging when high correlation orders must be kept in zero-field and solid-state NMR, corresponding to a state space scaling as a large polynomial of system size. Density matrix renormalization group (or tensor train) methods, which also exactly simulate density matrix time-evolution on classical computers, have alternatively been applied in the context of NMR [35], but have struggled to simulate large systems due to the irregular 3-dimensional interaction graphs that manifest between spins. Recent methodological developments, such as those discussed in Ref. [36], may be able to alleviate these issues to an extent, but it is yet unclear whether all simulation contexts will become tractable. Quantum Monte Carlo methods are another possible classical computing approach to simulating time-evolution of quantum spin systems. These methods sample dynamical trajectories rather than exactly simulating density matrix dynamics, and can, in principle, take advantage of decoherence in the NMR experiment and render the powder averaging overhead for solid-state NMR redundant similar to quantum hardware approaches. However, these methods remain largely unexplored in the context of NMR simulation. NMR thus provides a natural task where we can seek a practical quantum advantage from near-term quantum devices: simulation of noisy spin systems using noisy quantum computers.

## MATERIALS AND METHODS

NMR simulation circuits are run on a trapped ion quantum computer that uses the $^2S_{1/2}$ states of $^{171}Yb^+$ ions as the qubit states. We trap 15 ions in a chain for the simulation, and the circuits use 4 of those ionic qubits. Before each circuit iteration, ions are cooled using Doppler cooling and Raman sideband cooling, and then reset to the logical $|0\rangle$ state via optical pumping. The qubit state is manipulated using 355-nm pulsed Raman beams. Single qubit gates are implemented using SK1 pulses [37], and two-qubit gates are mediated by Mølmer-Sørensen interactions [38] —these gates are run sequentially. We measure the qubit states by shining 369-nm light resonant on the $^2S_{1/2} \rightarrow {}^2P_{1/2}$ cycling transition that scatters photons.

The time series data used to construct the NMR spectrum of acetonitrile was computed from 1000 shots of 102 different circuits, for each of which 8 different initial basis states were prepared. Each circuit of 1000 shots took approximately 60 seconds to run. The total runtime to collect the data was therefore approximately 13.5 hours. Most of the runtime was taken up by classical compilation overhead. Improving the gate compilation procedure would lead to a reduction of runtime in the near-term, thereby allowing faster computation of spectra.

While running circuits on the quantum machine, we perform system calibrations of trap voltages and gate amplitudes every hour to mitigate effects of system drift on circuit performance. We do not correct for state preparation and measurement (SPAM) errors in this study, and a table of our system's SPAM characterization is presented in Ref. [9].

## ACKNOWLEDGEMENTS

We thank Ilya Kuprov, Haribabu Arthanari, and Minh Tran for stimulating discussions.

**Funding:** K.S. and E.D. acknowledge funding from ARO grant number W911NF-20-1-0163, Harvard-MIT CUA, NSF EAGER-QAC-QCH award No. 2037687, and the NSF grant No. OAC-1934714. D.B., C.N., A.R., D.Z., O.K., M.C., and C.M. acknowledge support by the ARO through the IARPA LogiQ program (11IARPA1008), the NSF STAQ Program (PHY-1818914), the AFOSR MURIs on Dissipation Engineering in Open Quantum Systems (FA9550-19-1-0399) and Quantum Interactive Protocols for Quantum Computation (FA9550-18-1-0161), the ARO MURI on Modular Quantum Circuits (W911NF1610349), and the U.S. Department of Energy Quantum Systems Accelerator (QSA) Research Center (DE-FOA-0002253). S.C. acknowledges funding from the NSF GRFP. D.S. acknowledges funding from the Harvard Quantum Initiative Seed Funding program and AFOSR: Grant FA9550-21-1-0236. The Flatiron Institute is a division of the Simons Foundation.

**Author Contributions:** K.S. conceptualized the work, developed the circuit synthesis and compressed sensing methods, performed computer emulations, and analyzed the data. D.B., C.N., A.N, D.Z, O.K, M.C developed the experimental platform, performed the experiment, and collected the data. S.C. performed (classical computer) simulations to analyze the spurious spectral peak. D.S. analyzed the data, derived and performed resource estimates, and analyzed operator entanglement. E.D., C.M., and D.S. supervised the project. All authors contributed to writing the manuscript.

**Competing Interests:** All authors declare that they have no competing interests.

**Data Materials and Availability:** All data needed to evaluate the conclusions in the paper are present in the paper and/or the Supplementary Materials.

# SUPPLEMENTARY MATERIAL

**This PDF file includes:**

Figs. S1 - S9

Supplementary text

## S1. SCALING METHODS TO LARGER SYSTEMS

Here, we give details of the circuit synthesis and compressed sensing techniques used in the experimentally demonstrated quantum simulation, and then discuss how they may be applied to simulations of larger systems.

### A. Time-evolution circuit synthesis

We use the numerical optimization algorithm in Ref. [12] to synthesize the circuits implementing the time-evolution unitary $U(t) = \exp(-iHt/\hbar)$, with the Hamiltonian given in Eq. (S.1). The algorithm implements a bottom-up approach, building the single- and two-qubit gate decomposition of a $n$-qubit unitary by iteratively searching for a $m$-qubit gate decomposition with $m < n$. Initially, $m$ is set to $n-1$. The algorithm is hardware topology and gateset aware; to specialize for the trapped-ion system, we allow all-to-all connectivity of qubit interactions and choose Mølmer-Sørensen gates with variable angles as the interaction gate.

We choose a unitary error of $\epsilon = 10^{-2}$, with the synthesis algorithm producing a circuit in terms of Mølmer-Sørensen (MS) gates and generic single-qubit rotations that approximates the true time-evolution unitary within this error. We then iteratively perform a X-Z-X decomposition of each single-qubit rotations, commuting the trailing X rotation through each MS gate before decomposing the next single-qubit rotation. This optimization results in roughly two Z rotations and two X rotations after each MS gate. As Z rotations are implemented virtually in the trapped-ion system, the final circuit has only two physical single-qubit rotations for each MS gate, and thus the physical circuit depth is reduced compared to the initial output of the synthesis algorithm. An example of the final optimized circuit is shown in Fig. S1. Typically, the produced circuits were composed of up to 40 MS gates and 80 physical single-qubit gates.

### B. Compressed sensing

A general function in the frequency-domain that is nonzero in a specified frequency window can be reconstructed by Fourier transforming a corresponding time-domain signal that is uniformly sampled at the Nyquist rate. If the function is known to be sparse in the frequency domain, however, the time signal can be undersampled by choosing a non-uniform subset of time points which still capture the relevant information in the frequency domain [16]. The missing points on the original uniform time grid create artifacts in Fourier transform of the signal, however, which must then be removed using a compressed sensing reconstruction algorithm that exploits the assumed sparsity of the frequency signal. NMR spectra are often sparse as they are composed of a series of Lorentzian peaks, and therefore compressed sensing techniques allow for a dramatic reduction in the sampling required during an NMR experiment [11].

This sparsity can also be exploited in quantum simulations of NMR experiments by computing the FID at only the undersampled time points and then reconstructing the spectrum. We compute the FID at 102 out of the $N_s = 4096$ time points on the uniform grid, choosing the points according to a sine-weighted Poisson gap schedule. Such schedules have been shown to reduce undersampling artifacts [18]. The points are randomly chosen with the likelihood to pick a point $m+1$ on the uniform grid, given we have picked a point $m$, set by a Poission distribution with mean proportional to $\sin(\alpha\pi m/N_s)$. Specifically, we choose $\alpha = 0.5$, resulting in a schedule that is dense at short times before becoming increasingly sparse at later times. We find that this choice allows for much larger compression compared to a schedule with uniformly distributed gaps between points, or a schedule that is also dense at late times (corresponding to $\alpha = 1$). After computing the undersampled FID, we reconstruct the spectrum using the iterative soft thresholding (IST-S) algorithm [11].

In Fig. S2(a), we plot the FID computed via noisy emulations for all 4096 time points, and compare with the 102 points that were experimentally computed. This plot corresponds to Fig. 2(a) of the Main Text, but with the quantities depicted over the full time grid. In Fig. S2(b), we plot the spectrum computed after padding the experimental data with zeros for all time points that were not computed. We see that there is some signal at the spectral peaks expected from the noisy emulation, but the signal-to-noise is very large. The zero-padded spectrum in this plot corresponds to the green dots in Fig. 2(b) of the Main Text. In Fig. S2(c), we plot the compressed sensing reconstruction of the experimentally computed spectrum, and see that the signal-to-noise is dramatically improved. The reconstructed spectrum in this plot corresponds to the yellow curve in Fig. 2(c) of the Main Text.

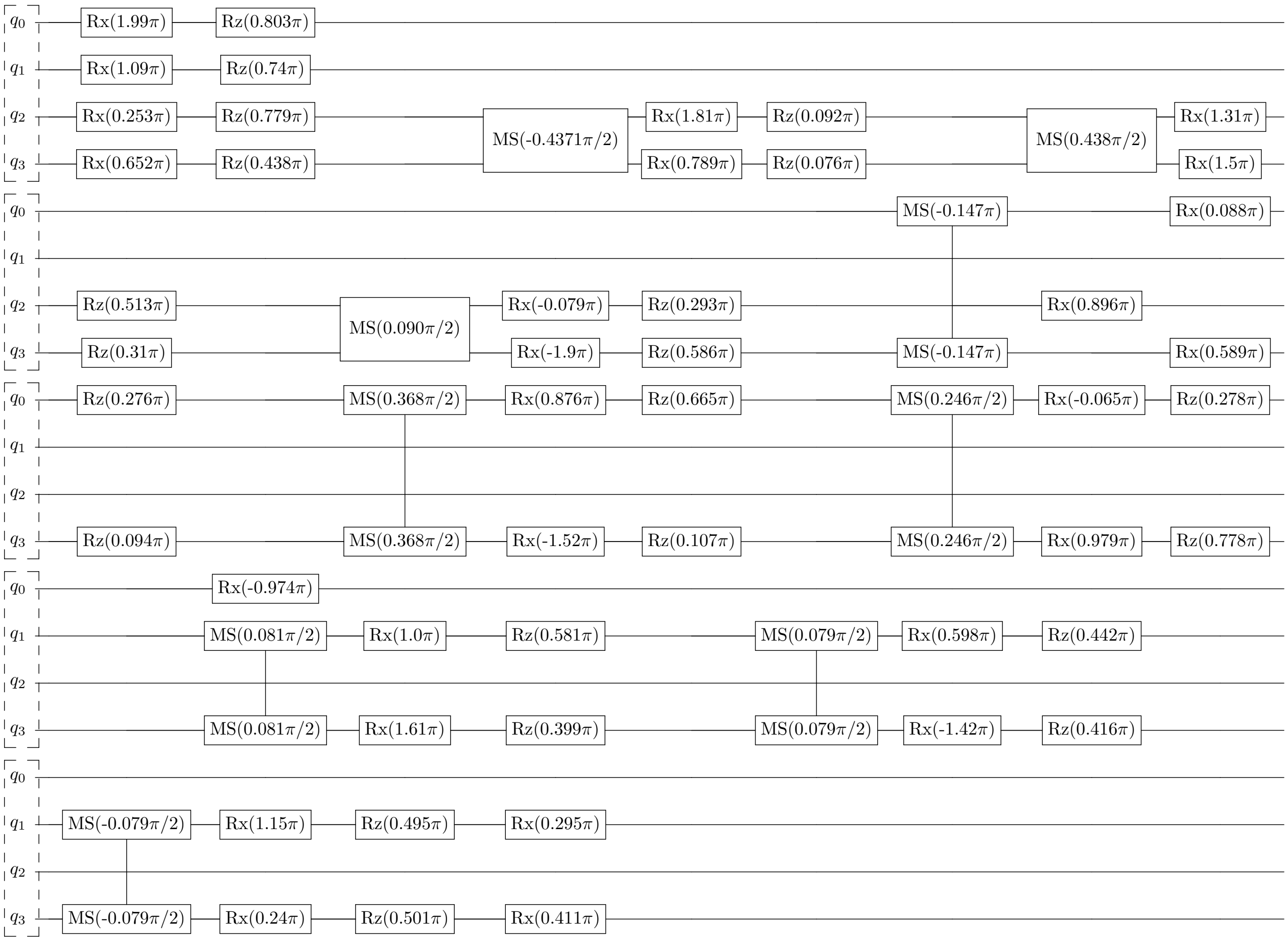

Figure S1: **Time-evolution circuit.** Example time-evolution circuit generated by numerical synthesis algorithm corresponding to $t = 0.07$ s. Circuit is split into five rows and read top to bottom, with the start of each row indicated by a dashed box around the four qubits in the experiment.

### C. Scaling to larger systems

The numerical optimization algorithm we use is likely to be limited to producing time-evolution circuits for systems of up to $\sim 7$ spins [12]. This tool can still prove useful, however, when scaling to large, classically-intractable NMR simulations by exploiting the cluster structure of these molecules (see Fig. 3A of the Main Text). The strongly-interacting clusters are usually formed from 4-7 spins, and the optimization algorithm can be used to synthesize the time-evolution circuit for each cluster. These circuits can then be combined with a Trotter formula to implement the time-evolution of the entire systems [24]. Compared to a Trotter decomposition of the entire system, such a hybrid approach can reduce the overall circuit depth, as discussed in Sec. S3 B. Furthermore, at the level of discretization estimated in Fig. 3 of the Main Text, the simulation times are small enough that the optimization should converge very quickly, potentially enabling real-time compilation of the overall time-evolution circuit.

We note that numerical circuit synthesis of small subsystems and compressed sensing techniques form a synergistic combination of tools. For example, cluster-exploiting Trotter formulas allow for an overall reduction in resource cost at all simulation times, while compressed sensing non-uniform sampling schedules may sample more densely from short times where the resource cost is smallest. On the hardware side, the all-to-all connectivity of trapped ions makes them well-suited to the interaction graphs within clusters, and may allow comparatively smaller gate counts for the cluster evolution circuits. The relatively slow cycle time of ion devices is ameliorated by compressed sensing techniques, which reduce the number of time points that must be sampled. The combination of numerical circuit synthesis, which exploits the clustered interaction structure of a system, and compressed sensing, which exploits sparsity of the observable of interest in the transform domain, may similarly prove useful for quantum simulations in quantum chemistry and condensed matter systems where both of these characteristics are often present [14].

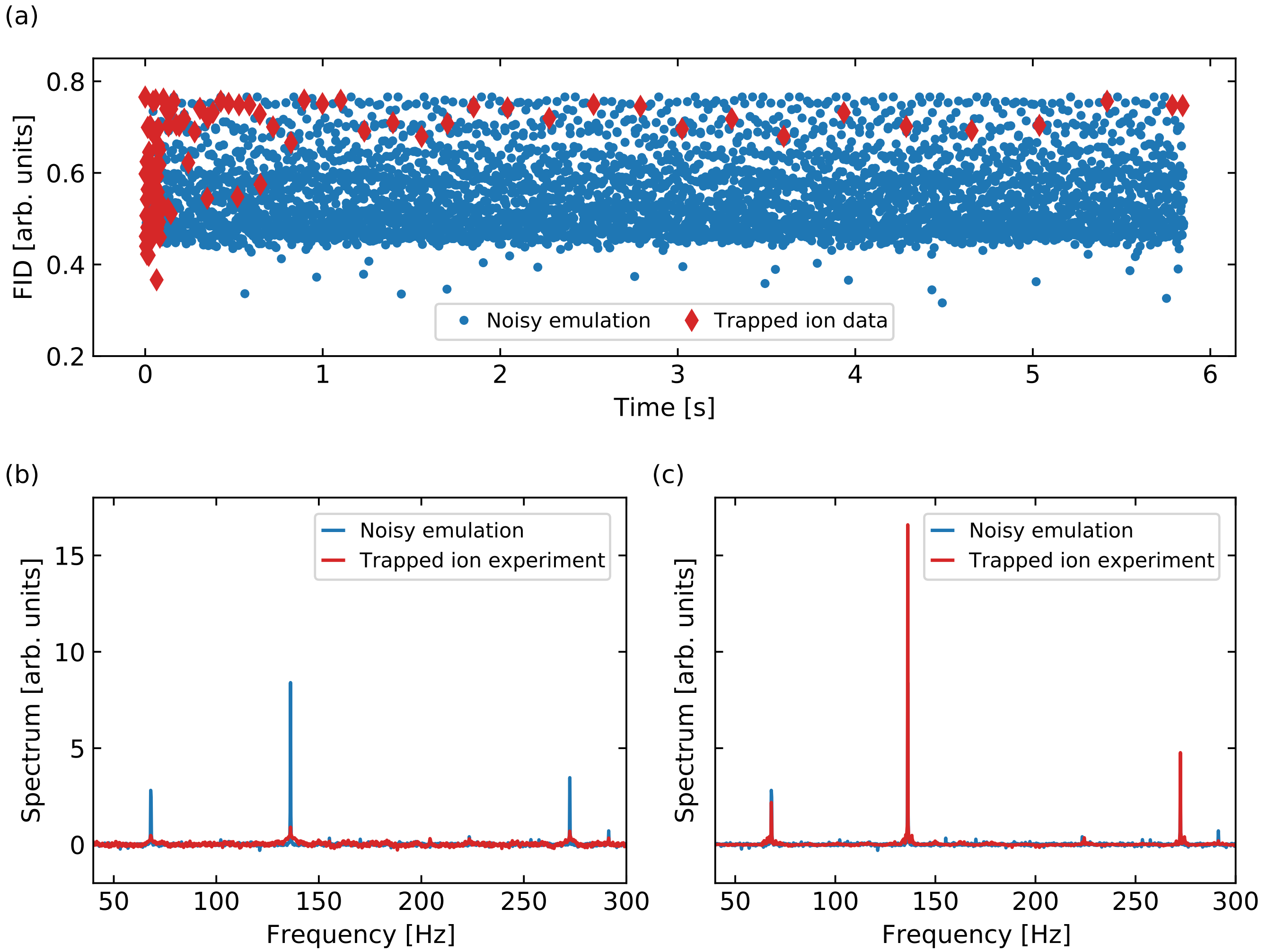

Figure S2: **Compressed sensing reconstruction.** **(a)** Comparison of the FID for noisy quantum circuit emulation on a fully sampled uniform time grid of 4096 points (blue circles) and the 99 data points experimentally measured on the ion trap device (red diamonds). **(b)** Fourier transform of the FID after replacing unsampled points with zeros. **(c)** Reconstructed spectrum after applying the iterative soft thresholding algorithm. The noise is modeled by two-qubit gates subject to both amplitude and phase damping with rates 0.005 and 0.035 respectively.

## S2. SPECTRAL PEAK AT J/2

The acetonitrile spectrum we compute on the trapped ion quantum computer, depicted in Fig. 1 of the main text, exhibits a resonance at frequency $J/2$ which does not appear in the NMR experiment of Ref. [10]. Here, we explain the origin of this additional peak and discuss how to prevent such artifacts from appearing in future experiments.

The zero-field nuclear spin Hamiltonian of acetonitrile is

$$\hat{H} = J\left(\hat{\mathbf{S}}_1 + \hat{\mathbf{S}}_2 + \hat{\mathbf{S}}_3\right)\cdot\hat{\mathbf{S}}_4, \tag{S.1}$$

where $\{\hat{\mathbf{S}}_1, \hat{\mathbf{S}}_2, \hat{\mathbf{S}}_3\}$ represent the three $^1$H and $\hat{\mathbf{S}}_4$ represents the $^{13}$C. The eight positive magnetization states used to

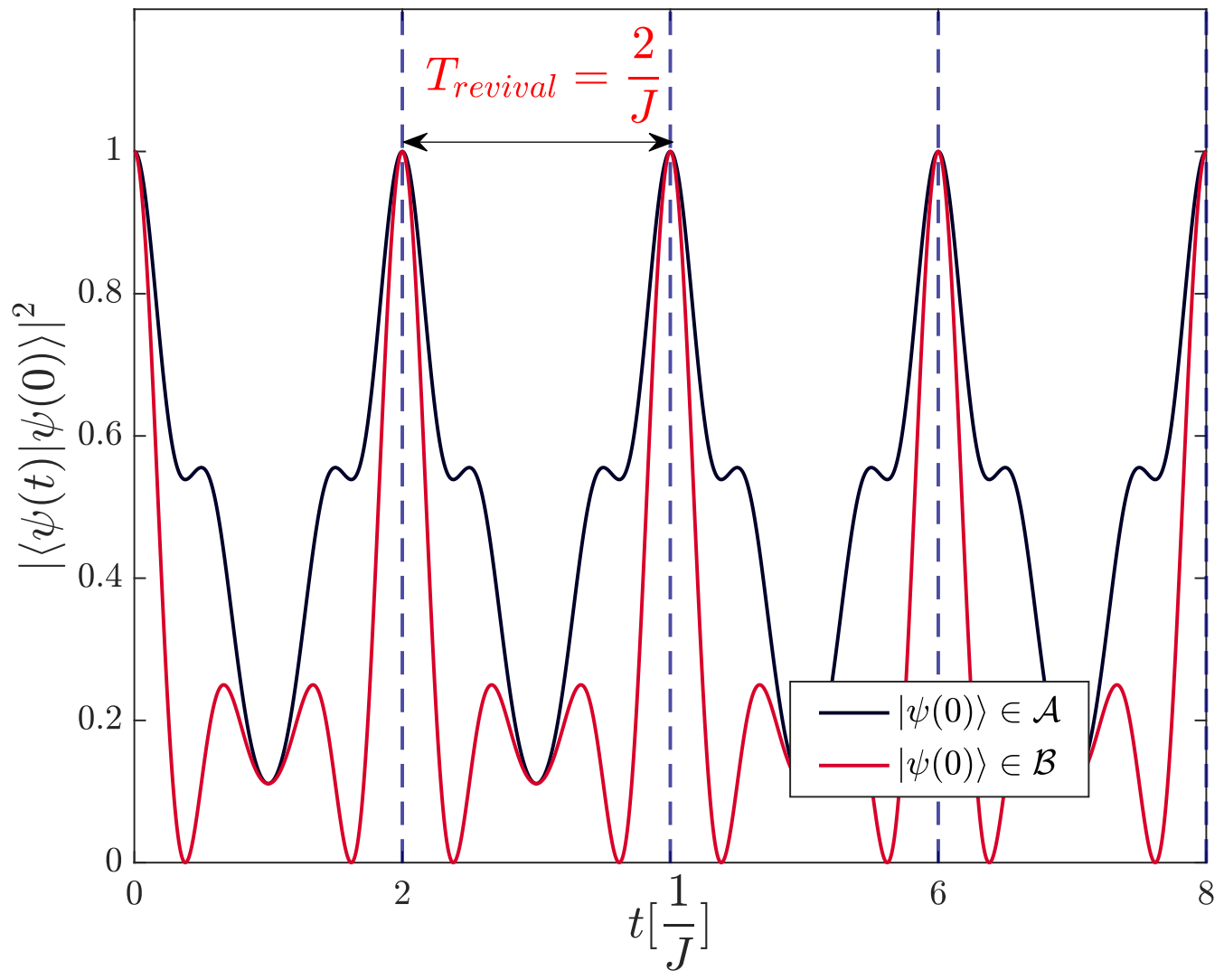

Figure S3: **Magnetization basis state revivals.** A system initialized in state $|\psi(0)\rangle$ selected from magnetization basis states $\mathcal{A} = \{\tilde{m}_1, \tilde{m}_3, \tilde{m}_5\}$ and $\mathcal{B} = \{\tilde{m}_4, \tilde{m}_6, \tilde{m}_8\}$ undergoes revivals with a period $2/J$.

compute the FID, see Eq. (3) of the main text, and their magnetizations are

$$
\begin{aligned}
|\tilde{m}_1 = 1.626\rangle &= |0000\rangle & \tilde{m}_1 &= 1.626 \\
|\tilde{m}_2 = 1.374\rangle &= |0001\rangle & \tilde{m}_2 &= 1.374 \\
|\tilde{m}_3 = 0.626\rangle &= |0010\rangle & \tilde{m}_3 &= 0.626 \\
|\tilde{m}_4 = 0.374\rangle &= |0011\rangle & \tilde{m}_4 &= 0.374 \\
|\tilde{m}_5 = 0.626\rangle &= |0100\rangle & \tilde{m}_5 &= 0.626 \\
|\tilde{m}_6 = 0.374\rangle &= |0101\rangle & \tilde{m}_6 &= 0.374 \\
|\tilde{m}_7 = 0.626\rangle &= |1000\rangle & \tilde{m}_7 &= 0.626 \\
|\tilde{m}_8 = 0.374\rangle &= |1001\rangle & \tilde{m}_8 &= 0.374
\end{aligned}
$$

The small, four spin Hilbert space of the NMR active nuclear spins of the molecule along with the molecules highly symmetric nature—as codified by the single interaction scale $J$ in Eq. (S.1)—combine to yield perfect revivals when the system is prepared in six of the above magnetization basis states. These states can be grouped into the triads $\mathcal{A} = \{\tilde{m}_1, \tilde{m}_3, \tilde{m}_5\}$ and $\mathcal{B} = \{\tilde{m}_4, \tilde{m}_6, \tilde{m}_8\}$ and we depict their revivals in Fig. S3. When viewed in the energy eigenstate basis, each of these six magnetization basis states only has weight on energy eigenstates whose eigenvalues are integer multiples of $J/2$. Consequently, all energies relevant to the dynamics are commensurate with a smallest splitting of $J/2$, which leads to the state reviving with perfect fidelity at this frequency. The revival of each state is mirrored in its entanglement dynamics. Each magnetization state begins in an unentangled product state, non-monotonically accrues entanglement over a period $T = \frac{2}{J}$, and then dis-entangles as it returns to the original product state. The high symmetry and small size of the molecule therefore causes the dynamics to defy usual expectations of ergodicity, with the entanglement of a system initially prepared in one of the states in $\mathcal{A}$ or $\mathcal{B}$ oscillating at a frequency $J/2$ instead of growing monotonically in time.

The numerical optimization algorithm, Ref. [12], we use to synthesize time-evolution circuits for each time point reflects this oscillating entanglement in the gate depth of the synthesized circuits. Specifically, times at which the system is more heavily entangled correspond to deeper circuits with a larger number of two-qubit gates, as can be seen in Fig. S4.

Noise in the system affects deeper circuits more than shallower ones, and therefore imprints the $J/2$ entanglement oscillation onto the experimentally measured observable by lowering the fidelity of the signal at this frequency. We can gain visibility into this process by computing the average Bhattacharyya coefficient (BC) between the measured basis state populations and noiseless emulations of the circuits for each time point. The BC is defined as

$$\mathrm{BC} = 1 - \frac{1}{2}\sum_j \left(\sqrt{p(j)} - \sqrt{q(j)}\right)^2 , \tag{S.2}$$

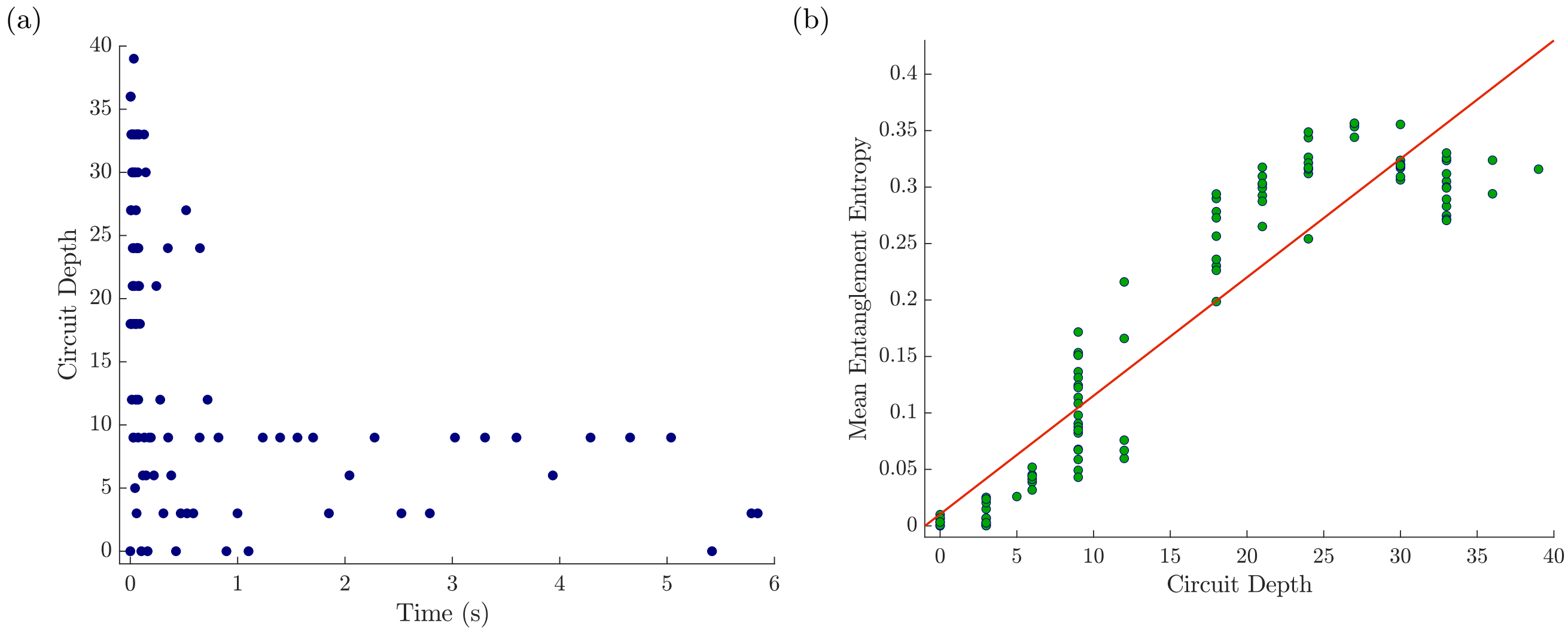

Figure S4: **Synthesized circuit depth and entanglement. (a)** Two-qubit gate count of synthesized time-evolution circuits for each FID evolution time measured on the trapped-ion device. **(b)** Entanglement entropy of the system at a particular evolution time compared to the two-qubit gate count of the circuit implementing that evolution. We average the final entanglement entropy for systems initialized in each of the eight magnetization basis states used to compute the FID.

where $j$ runs over all computational basis states, and $p(j)$ and $q(j)$ are the corresponding occupation probabilities, given by the diagonal elements of the density matrix, of the two states beings compared. The BC gives a measure of the fidelity of the experimental runs and we plot it for every experimentally measured time point in Fig. S5(a). We see that it varies as a function of time, and these oscillations correspond to time-evolution circuits that have a larger two-qubit gate count as shown in Fig. 2C of the Main Text. In Fig. S5(b), we use the same compressed sensing algorithm used to compute the NMR spectrum to reconstruct the Fourier transform of the the BC dynamics depicted in Fig. S5(a). We observe a sharp peak at $J/2$, confirming that the fidelity of the experimental oscillates at a frequency of $J/2$. These oscillations corresponding to oscillating depths of the synthesized circuits, which in turn reflect the entanglement revivals of the molecule's underlying dynamics. The above story confirms why noisy circuit simulations, such as the one depicted in Fig. 2A of the main text, also exhibit this $J/2$ resonance peak in the computed NMR spectrum regardless of the type of decoherence channel used to simulate noise.

artifact peaks such as the $J/2$ resonance can easily be removed in future experiments. By padding all time-evolution circuits so that they have roughly the same depth as the deepest synthesized circuit, the noise in the system can no longer imprint any frequency on the measured signal as the gate depths no longer oscillate. We show in Fig. S6 that such padding dramatically decreases the height of the $J/2$ peak in noisy circuit simulations of the experiment, where noise is modeled as a depolarizing channel on the gates. The padding will, however, slightly decrease the overall fidelity of the computed FID as every point will be subject to as much noise as the deepest time-evolution circuit. If we desire to compute the maximum fidelity signal allowable by hardware, we can compute FID twice - once with padded circuits and once without. Any feature that vanishes in the padded experiment can be removed from the higher fidelity non-padded experiment.

Lastly, we note that artifact peaks are unlikely to appear during quantum simulations of the majority of NMR experiments, and become increasingly unlikely when scaling to classically intractable systems. Small molecules which do not exhibit the high degree of symmetry exhibited in Eq. (S.1) are unlikely to exhibit the dynamical revivals at the heart of artifact peaks. Larger systems, including those with some symmetry, are even less likely to exhibit revivals as entanglement spreads throughout the system. In fact, classically intractable systems are intractable precisely because quantum correlations spread quickly throughout the system. Furthermore, quantum simulation algorithms that generalize to larger systems, such as product formulas, typically have gate depths that monotonically increase with the simulation time. This relationship is also directly true for analog quantum simulation. Noise in these cases will lead only to a broadening of spectral peaks, and cannot imprint any artifact frequencies on the measured signal. Therefore, artifact peaks are unlikely to be a common concern during quantum simulations of NMR experiments. Even in small, highly symmetric systems where the peaks might appear, circuit padding is a cheap way to remove these artifacts.

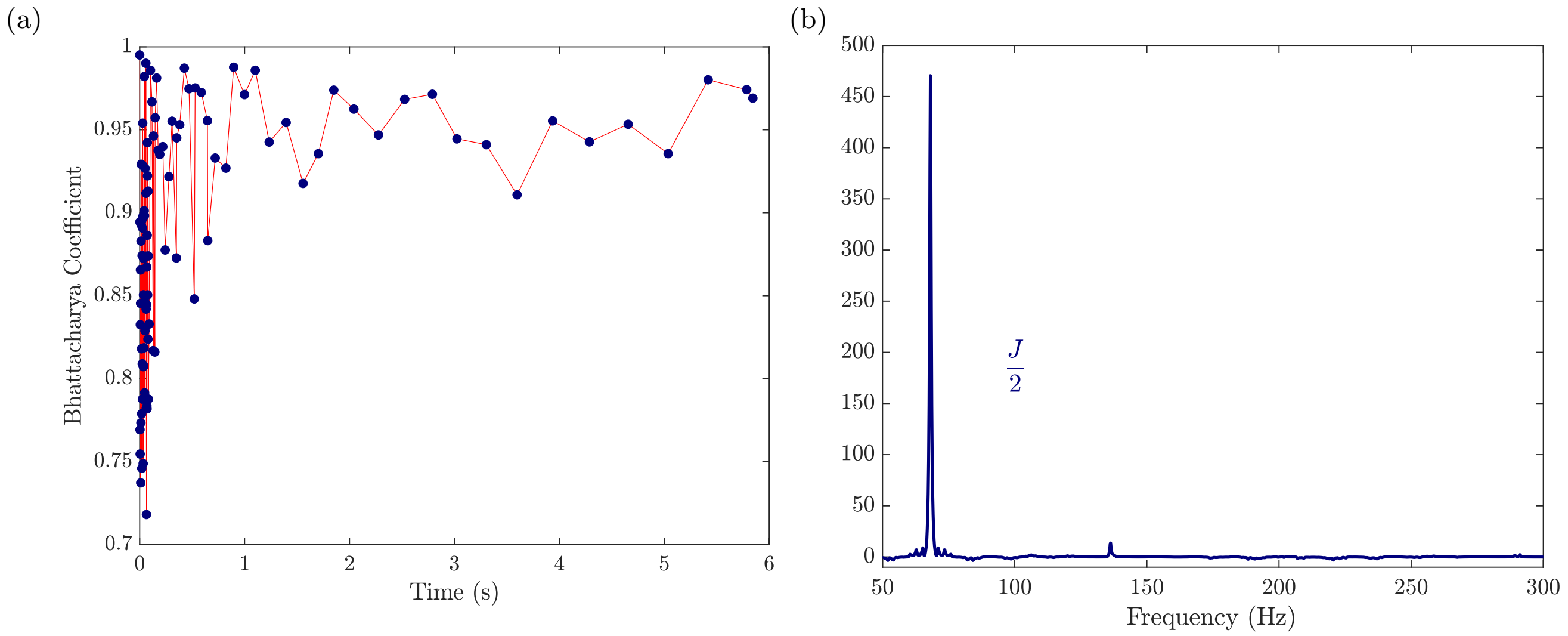

Figure S5: **Bhattacharyya coefficient between trapped ion measurements and noiseless emulation of the experiment.** **(a)** BC vs evolution time. **(b)** Compressed sensing reconstruction of the frequency spectrum of the BC. We see that the BC, a measure of the fidelity of the system, only varies at the frequency $J/2$ with which the system's entanglement, and therefore circuit depth, oscillates.

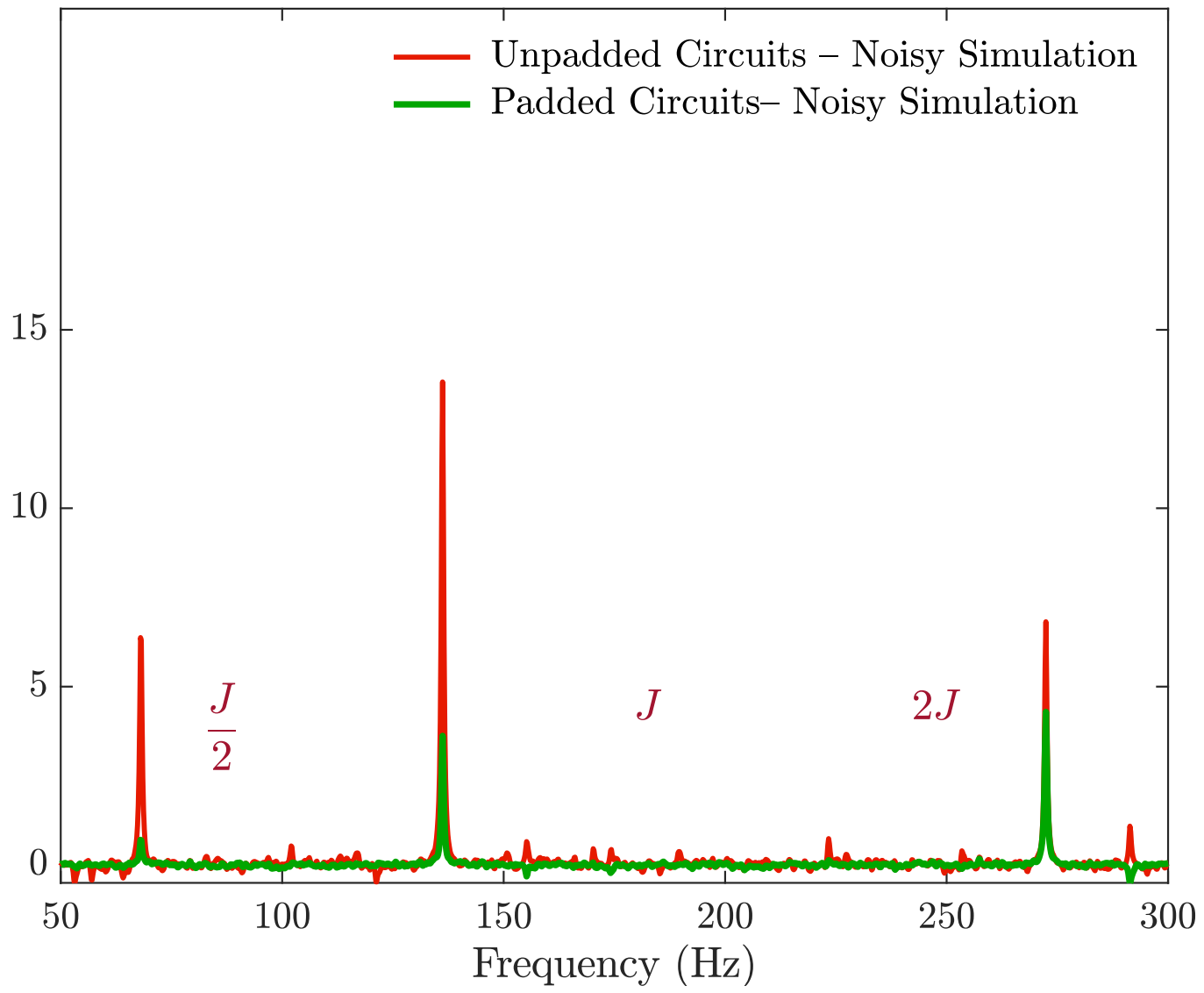

Figure S6: **Noisy circuit simulation.** The zero-field NMR spectrum of acetonitrile computed using noisy circuit simulations with and without padding. The padded circuits no longer have depths that oscillate in accordance with the system's entanglement, and therefore do not exhibit the artifact peak at $J/2$. Noise was simulated by by adding a depolarizing channel to each gate, with a rate of $10^{-3}$ for single-qubit gates and $10^{-2}$ for two-qubit gates.

## S3. RESOURCE ESTIMATES FOR NMR SIMULATION

The standard Hamiltonian simulation task (e.g. by Suzuki-Trotter product formula, hereafter product formula), seeks to approximate exact unitary dynamics within a finite precision $\epsilon$ (clarified below). However, the spectral resolution—alternatively, the line-width—of NMR experiments are set by the dephasing of the nuclear spins. The inherent dephasing in the experiments we seek to simulate reduces the resource cost simulation compared to purely coherent systems. In what follows, we show how this discrepancy between the standard Hamiltonian simulation task (simulating a unitary to finite *precision*) and the task of employing Hamiltonian simulation as a sub-routine to compute an NMR spectrum (with finite *spectral resolution*) can yield gate counts that are several orders of magnitude smaller for the latter task, making it tractable on NISQ devices. To perform Hamiltonian simulation, we proceed by using a variant of first-order product formula that exploits the clustered structural motif present in many molecules that are classical challenging to simulate. In what follows, we first elucidate the distinction between the standard Hamiltonian simulation task and the task at hand and provide bounds on the requisite two-qubit gate fidelity and gate counts for computing NMR spectra with finite resolution. We round out our discussion by providing commutator bounds relevant to estimating the quantum resources required to simulate the NMR spectra of molecules with a clustered interaction structure.

### A. Approximating Hamiltonian Dynamics for NMR Simulations

We begin by clarifying the difference between a standard Hamiltonian simulation task and the task of using Hamiltonian simulation to simulate NMR experiments. We consider performing Hamiltonian simulation via first order Trotterization, a simple but powerful and gate efficient method for simulating Hamiltonian dynamics. Keeping our discussion somewhat general for the moment, let's say we have a Hamiltonian $H = \sum_\mu h_\mu$, composed out of a number of $N_c$ terms $h_\mu$. We'd like to replace the time-evolution operator $U = e^{-i\Delta t H}$ , for some small time-step $\Delta t$ by our simple product formula:

$$\tilde{U} = \prod_{\mu=1}^{N_c} e^{-i\Delta t h_\mu}.$$

It follows from Baker-Campbell-Hausdorff (BCH), by keeping only lowest order contributions in $\Delta t$, that

$$\|U - \tilde{U}\| \leq \frac{(\Delta t)^2}{2} \sum_{\mu=1}^{N_c} \| \sum_{\nu>\mu} [h_\nu, h_\mu]\|, \tag{S.3}$$

as obtained in Ref. [7]. Consequently, we could also write the fidelity of the simulation as

$$\mathcal{F}_{\delta t} = \|U\tilde{U}\| \geq e^{-\beta(\Delta t)^2/2}, \quad \text{where} \quad \beta = \sum_{\mu=1}^{N_c} \| \sum_{\nu>\mu} [h_\nu, h_\mu]\| \tag{S.4}$$

Let's say we'd like to evolve for a total time $T = r\Delta t$, then the total fidelity will be

$$\mathcal{F}_{PF}(T) = \prod_{i=1}^{r} \mathcal{F}_{\delta t_i} = e^{-r\beta(\delta t)^2/2} = \exp\left(-\frac{\beta T^2}{2r}\right), \tag{S.5}$$

If we want a precision $\epsilon$, then we need $\mathcal{F} = 1 - \epsilon \approx e^{-\epsilon}$. This sets the Trotter number $r$ to achieve a precision $\epsilon$:

$$r_\epsilon = \frac{\beta T^2}{2\epsilon}. \tag{S.6}$$

Achieving this can be quite challenging for even intermediate times, in particular in NISQ settings.

The task of simulating the relevant dynamics corresponding to an NMR experiment does not, however, require the approximation of unitary dynamics generated by the Hamiltonian to finite, time-independent precision $\epsilon$. It requires instead the simulation of a spectrum to finite spectral resolution $\Delta f$. In an NMR experiment, $\Delta f \sim \gamma$, the dephasing rate of a single nuclear spin in experimentally interrogated sample. Thus, the task of simulating an NMR experiment with resolution $\Delta f$ is equivalent to simulating Hamiltonian dynamics of a sample of $N$ spins in which each spin decoherences independently with an *effective* dephasing rate $\gamma = \frac{\Delta f}{2\pi}$. Such dephasing exponentially degrades the fidelity, vis-a-vis perfect unitary dynamics given by the Hamiltonian, as $\mathcal{F}_{NMR} \sim e^{-\gamma N t}$. Thus, there is a subtle but essential distinction between the task of approximating a unitary to multiplicative error $\epsilon$ and performing Hamiltonian simulation to compute a spectrum with finite spectral resolution $\Delta f$. If, for the moment, we neglect decoherence in our quantum hardware and consider only algorthmic error due to our product formula, as given in S.5, then setting $\mathcal{F}_{PF}(T) = \mathcal{F}_{NMR}(T)$ we find a upper bound for the minimal necessary Trotter number:

$$r_{NMR} = \frac{\beta T}{2\gamma N}. \tag{S.7}$$

As typical experiments interrogate regimes up to where $\gamma T \sim 1$, the number of Trotter steps $r$ is reduced by a factor of $\frac{N}{\epsilon}$ as compared to the case of fixed precision—for the examples examined in the Main Text, this corresponds to a $O(10^3 - 10^4)$-fold decrease in the number of steps required to realize the longest dynamics relevant to experiment.

For a more careful estimate of the necessary resources, we also account for the decrease in fidelity due to decoherence in our quantum hardware. We describe the decaying fidelity of our experiment due to hardware error as $\mathcal{F}_h(T) \sim F^{rN_g}$, where $F$ is the two-qubit gate fidelity, $r$ is the Trotter number, and $N_g$ is the number of two-qubit gates required to realize a particular Trotter step. Note that by such a description, we assume that hardware error and algorithmic error are independent of each other and that two-qubit gates dominate the hardware error. The product of $\mathcal{F}_h(T)$ and $\mathcal{F}_{PF}T$ must be greater than or equal to $\mathcal{F}_{NMR}$ in order to perform reliable simulation. We obtain, the following requirement:

$$- N_g \log(F) + \beta(\Delta t)^2/2 \leq \gamma N \Delta t. \tag{S.8}$$

This will have a solution for the Trotter time step $\Delta t$ as long as

$$F \geq e^{-\gamma N \Delta t / N_g}. \tag{S.9}$$

Requiring our quantum simulation fidelity to match the fidelity of the experiment —thereby recasting S.8 as an equality —and re-arranging, we can establish an equation for $\gamma \sim \Delta f$:

$$\gamma = \frac{1}{N}\left(\frac{\Delta t \beta}{2} - \frac{N_g \log(F)}{\Delta t}\right) \tag{S.10}$$

By optimizing over the Trotter step $\Delta t$, we can set the ultimate resolution $\Delta f_{opt}$ of our experiment:

$$\Delta f_{opt} = \frac{\gamma_{opt}}{2\pi} = \frac{1}{2\pi N}\sqrt{2N_g \beta \log(1/F)} \tag{S.11}$$

Note that this optimal resolution is a simple function of the fidelity of the quantum hardware employed, given by $F$, and the efficiency of the algorithm used, as encoded by $\beta$ and $N_g$. On the hardware side, improving gates and thereby improving the gate fidelity, parametrized by $F$, would lower the resolution. Similarly, on the algorithmic side, finding more efficient circuits to realize a single Trotter step (lowering $N_g$) or better product formulae (lowering $\beta$), would improve the resolution—in what follows, we provide strategies on how to achieve both of these algorithmic improvements.

### B. Commutator Bounds for Clustered Hamiltonians

As implied by Eq. (S.11), decreasing $\beta$ (defined above in Eq. (S.4)) small would improve the resolution of the simulation. In what follows, we compute $\beta$, assuming an NMR Heisenberg Hamiltonian with clustered interactions, first in the standard way and then by taking advantage of the cluster motif. We show that by doing the latter, we can reduce $\beta$ substantially.

Before beginning this program, it is useful to establish some intuition for how clustered interactions reduce $\beta$ vis-a-vis the case of all-to-all couplings. If we take an all-to-all model with some typical coupling $J/\sqrt{N}$ (taking into account Kac normalization to keep the energy extensive in the system size $N$), we find

$$\beta = O(J^2 N^2), \quad \text{such that} \quad r \sim \left(\frac{JN}{\gamma}\right)^2 (\gamma T). \tag{S.12}$$

The situation changes for clustered Hamiltonians where each spin typically interacts with a sub-extensive number of spins $k$. For each term in the latter, only $k$ terms contribute in the commutator and there are $kN$ terms, yielding:

$$\beta \sim k^2 N J^2, \quad \text{such that} \quad r \sim \left(\frac{Jk}{\gamma}\right)^2 (\gamma T). \tag{S.13}$$

Note that, due to the clustered nature of the interactions, $r$ does not scale with $N$: Increasing the number of spins does not increase the Trotter number.

Having established an heuristic derivation for the scaling of $\beta$ and thereby the Trotter number, we turn to the present situation of a Heisenberg model with local fields.

$$H = \sum_{ij} J_{ij}(S_i^x S_j^x + S_i^y S_j^y + S_i^z S_j^z) + \sum_i h_i S_i^z. \tag{S.14}$$

where all terms can be labeled by $\mu = (i, j, \sigma)$, indicating the bond and the operator that is acted with. If we perform a scheme which alternates all Ising-$XX$ gates with all Ising-$YY$s and all Ising-$ZZ$s, then $\beta$ is comprised of three terms, defined below: $\beta_1$, $\beta_2$, $\beta_3$.

We first bound $\beta_1$, given by:

$$\beta_1 = \sum_{ij} |J_{ij}| \| \sum_{kl} J_{kl} [S_k^y S_l^y + S_k^x S_l^x, S_i^z S_j^z] \|, \tag{S.15}$$

Straightforward algebra brings us to

$$\beta_1 = \sum_{ij} |J_{ij}| \| \sum_k \left( 2J_{ki}(S_k^x S_i^y - S_k^y S_i^x) S_j^z + 2J_{kj}(S_k^x S_j^y - S_k^y S_j^x) S_i^z \| \right). \tag{S.16}$$

where the fact that $J_{ij} = J_{ji}$ is used. We therefore have:

$$\beta_1 \leq \sum_{ij} 2|J_{ij}| \sum_k \left( \| J_{ki}(S_k^x S_i^y - S_k^y S_i^x) S_j^z + J_{kj}(S_k^x S_j^y - S_k^y S_j^x) S_i^z \| \right), \tag{S.17}$$

which can again be bounded as:

$$\beta_1 \leq 2 \sum_{ij} |J_{ij}| \sum_k \left| J_{ki} \right| \| (S_k^x S_i^y - S_k^y S_i^x) S_j^z \| + |J_{kj}| \| (S_k^x S_j^y - S_k^y S_j^x) S_i^z \| \right). \tag{S.18}$$

Computing the norm directly—$\|(S_k^x S_i^y - S_k^y S_i^x) S_j^z\| = 1/4$ —we arrive at:

$$\beta_1 \leq \sum_{ijk} |J_{ik}||J_{ki}| \tag{S.19}$$

We can similarly bound $\beta_2$, which is given by:

$$\beta_2 = \sum_{ij} |J_{ij}| \| \sum_{kl} J_{kl} [S_k^y S_l^y, S_i^x S_j^x] \|, \tag{S.20}$$

as

$$\beta_2 \leq \frac{1}{2} \sum_{ijk} |J_{ik}||J_{kj}| \tag{S.21}$$

Finally there are local field terms, which if we do them together with the ZZ gates would give

$$\beta_3 = \sum_i |h_i| \| \sum_{kl} J_{kl} [S_k^y S_l^y + S_k^x S_l^x, S_i^z] \|, \tag{S.22}$$

which gives

$$\beta_3 = 2 \sum_i |h_i| \| \sum_k J_{ki} (S_k^x S_i^y - S_k^y S_i^x) \|, \tag{S.23}$$

and thus

$$\beta_3 \leq \sum_{ij} |h_i||J_{i,j}| \tag{S.24}$$

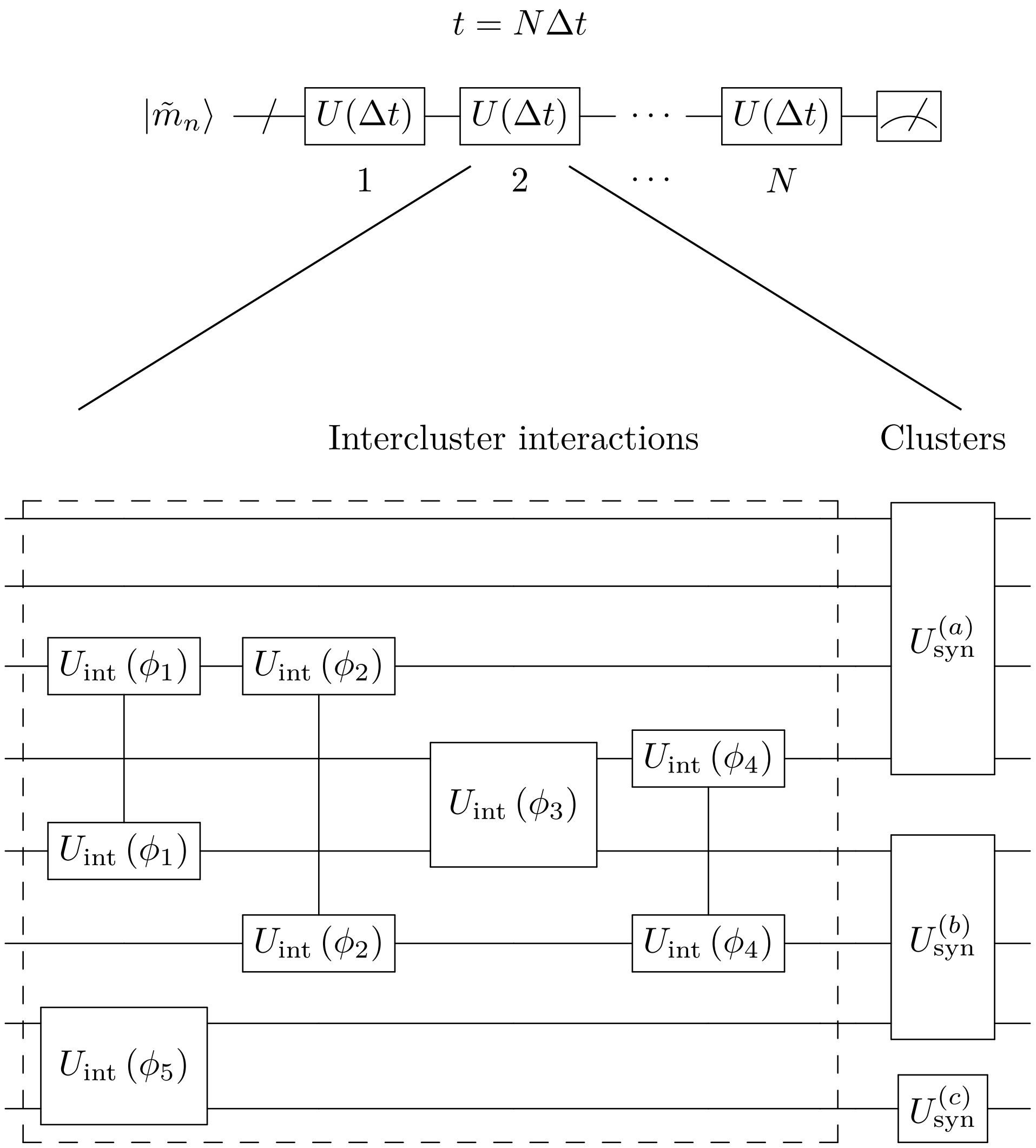

Figure S7: **Cluster Trotter circuit.** Schematic of quantum circuits simulating the time-evolution of challenging NMR systems. The total time-evolution is split into $N$ identical increments. Each increment is composed of numerically synthesized circuits enacting the time-evolution of strongly interacting clusters of spins, along with two-qubit gates enacting interactions between spins from different clusters. An example of a numerically synthesized circuit is given in Fig. S1.

However, one can obtain a tighter bound by treating the Z-gates as a single global gate. Doing so yields:

$$\tilde{\beta}_3 = \| \sum_{kl} J_{kl}[S_k^y S_l^y + S_k^x S_l^x, \sum_i h_i S_i^z]\|, \tag{S.25}$$

in which case we'd arrive at

$$\tilde{\beta}_3 \leq \frac{1}{2} \sum_{ij} |h_i - h_j||J_{i,j}| \leq \beta_3. \tag{S.26}$$

Therefore, we have that the total spectral norm of the nested commutator $\beta$ is bounded by:

$$\beta \leq \beta_1 + \beta_2 + \tilde{\beta}_3 \leq \frac{3}{2} \sum_{ijk} |J_{ik}||J_{kj}| + \frac{1}{2} \sum_{ij} |h_i - h_j||J_{i,j}|. \tag{S.27}$$

For many molecules, the interactions between nuclear spins follows a clustered motif. In particular, we examine the computational resources to simulate classically hard molecules that are composed of strongly-interacting clusters tethered together with weakly interacting links. We leverage the clustered motif by performing a variant of first-order Trotter formula wherein the dynamics of the small clusters are numerically synthesized to high precision, using for example the algorithm of Ref. [12], while weak interactions between such clusters are rendered via a pairwise Trotter decomposition of the Hamiltonian. A quantum circuit schematic depicting such a cluster-exploiting Trotter decomposition is given in Fig. S7.

To understand how this modifies $\beta$, we consider the following Hamiltonian:

$$H = H_c + \sum_{i,j} J_{i,j}(S_i^x S_j^{x,c} + S_i^y S_j^{y,c} + S_i^z S_j^{z,c}) + \sum_i h_i S_i^z, \tag{S.28}$$

and

$$H_c = \sum_{k,l} V_{k,l}(S_k^{x,c} S_k^{x,c} + S_k^{y,c} S_l^{y,c} + S_k^{z,c} S_l^{z,c}) + \sum_k h_k S_k^{z,c}, \tag{S.29}$$

where the superscript simply indicates that the operator belongs to the cluster $H_c$. If we could synthesize the Hamiltonian of the cluster efficiently (see below), we find:

$$\beta \leq \frac{3}{2}\sum_{i,j,k} |J_{i,k}V_{k,j}| + \frac{3}{2}\sum_{i,j,k} |J_{i,k}J_{k,j}| + \frac{1}{2}\sum_{i,k} |J_{i,k}||h_k - h_i|. \tag{S.30}$$

By comparing Eq. (S.30) with Eq. (S.27), we see that the latter avoids terms with intra-cluster couplings (i.e. terms like $V_{i,j}V_{k,l}$). For the cases considered in the Main Text, making use of the cluster structure in this manner reduces the Trotter number by one to two orders of magnitude.

We note that both the complexity of the coupling graph as well as number of coupling terms that determine the Trotter error, thus allowing for cases where a molecule with a simpler coupling graph may have less error than a slightly smaller molecule with a more complicated coupling graph. Indeed, this is seen in Fig. 3 of the Main Text.

## S4. CORRELATION SPREADING FROM NMR HAMILTONIANS

While NMR was originally proposed as a platform for quantum computing, the association of quantum speedup with quantum entanglement has raised questions about the possibility of speedup in NMR based quantum computing. It has been shown that the state of the system is always separable [39], severely restricting the accessible phase space. While large entanglement in pure states precludes efficient classical simulation, large operator entanglement can do the same for mixed states if one has access to two-time correlation function. This is exactly the setting of standard NMR spectroscopy. As far as we know, the argument was first laid out in [40] and called strong correlations, for lack of a better word. For completeness, we repeat the arguments of Ref. [40] for our specific setup, i.e. we have computed the operator entanglement entropy and Schmidt rank for the $S^z_{\text{tot}}$-magnetization evolving under a few relevant Hamiltonians, including two real molecules and some random matrices, see Fig. S8.

In complete analogy with the entanglement entropy for a state, we can partition are system in two subsystems $A$ and $B$, and define a complete set of orthonormal basis operators on each subsystem $A_i$ and $B_i$ respectively. Any operator can than be decomposed into

$$O = \sum_i \lambda_i A_i \otimes B_i. \tag{S.31}$$

We define the operator entanglement as

$$S = -\sum_i |\lambda_i|^2 \log_2(|\lambda_i|^2). \tag{S.32}$$

The initial magnetization

$$S^z_{\text{tot}} = \sum_i S_i^z = \sum_{i\in A} S_i^z \otimes \mathbb{I}_B + \sum_{i\in B} S_i^z \otimes \mathbb{I}_A, \tag{S.33}$$

has Schmidt rank 2 with respect to any bi-partition of the system. As shown in Fig. S8 the operator entanglement, in realistic systems, grows fast when this operator is evolved unitarily in time. It saturates at a value which is close to the maximal allowed entropy. The large, and ultimately extensive (see Fig. S9), operator entanglement implies that the unitary dynamics becomes classically intractable. While this highlights the complexity of quantum dynamical systems at infinite temperature, it should be noted that real systems aren't isolated and the environment induced decoherence tends to destroy the operator entanglement structure. Whether or not the particular evolution is classicaly hard depends on the balance between the entanglement growth and the decoherence, e.g. recent development in tensor networks have lead to the possibility of simulating noisy quantum systems when the noise levels exceed a minimal threshold [41].

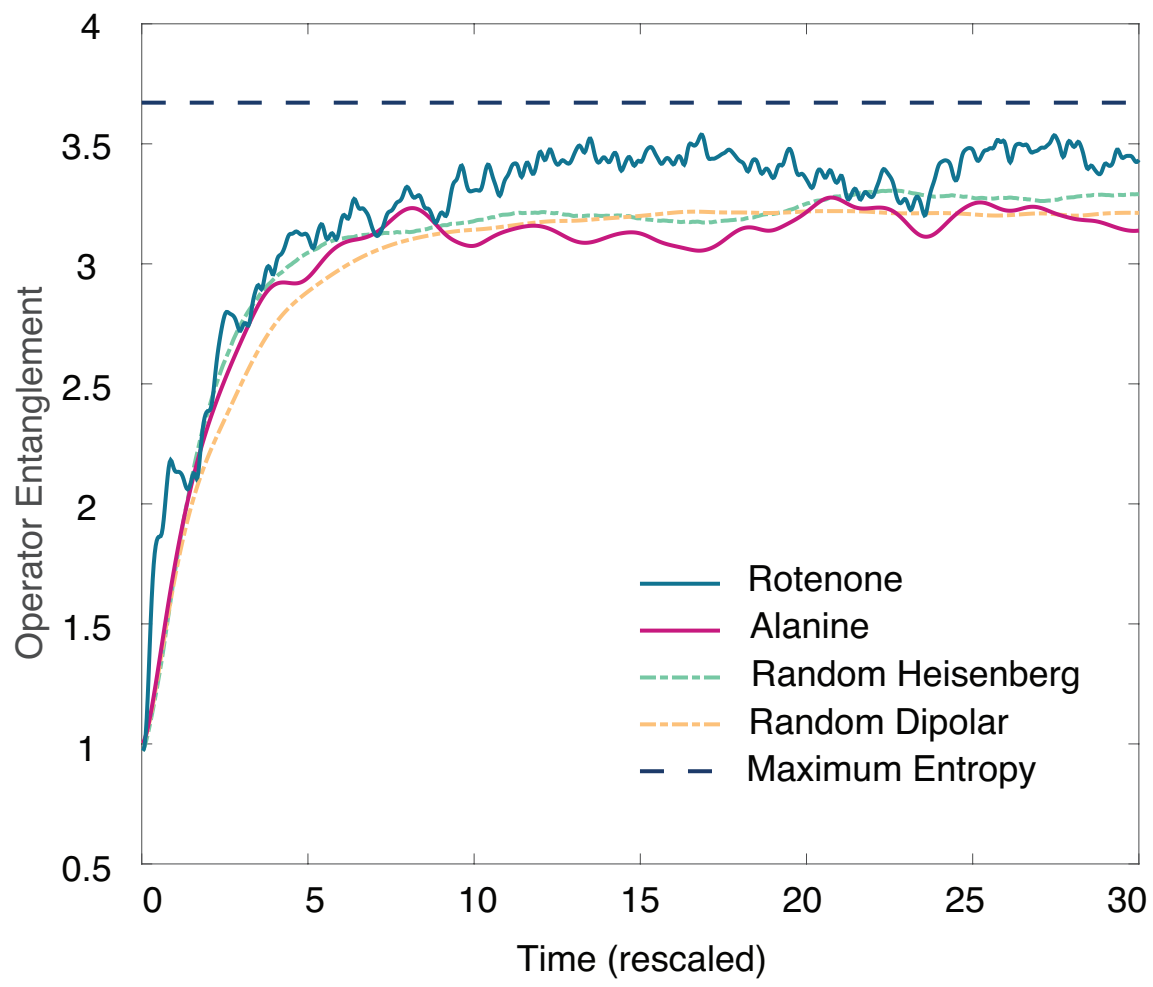

Figure S8: **Operator Entanglement I.** After a quench the operator entanglement in the total magnetization grows from its initial value $S = 1$ (Schmidt rank 2) to a value that is close to the maximal allowed entanglement. We show the evolution of two different molecules during an NMR spectroscopy sequence. For comparison we also show the operator entanglement in the corresponding random Hamiltonians, i.e. Hamiltonians where the couplings between the nuclear spins are drown from a Gaussian random matrix. The results are comparable, indicating fast spreading of correlations during NRM spectroscopy.

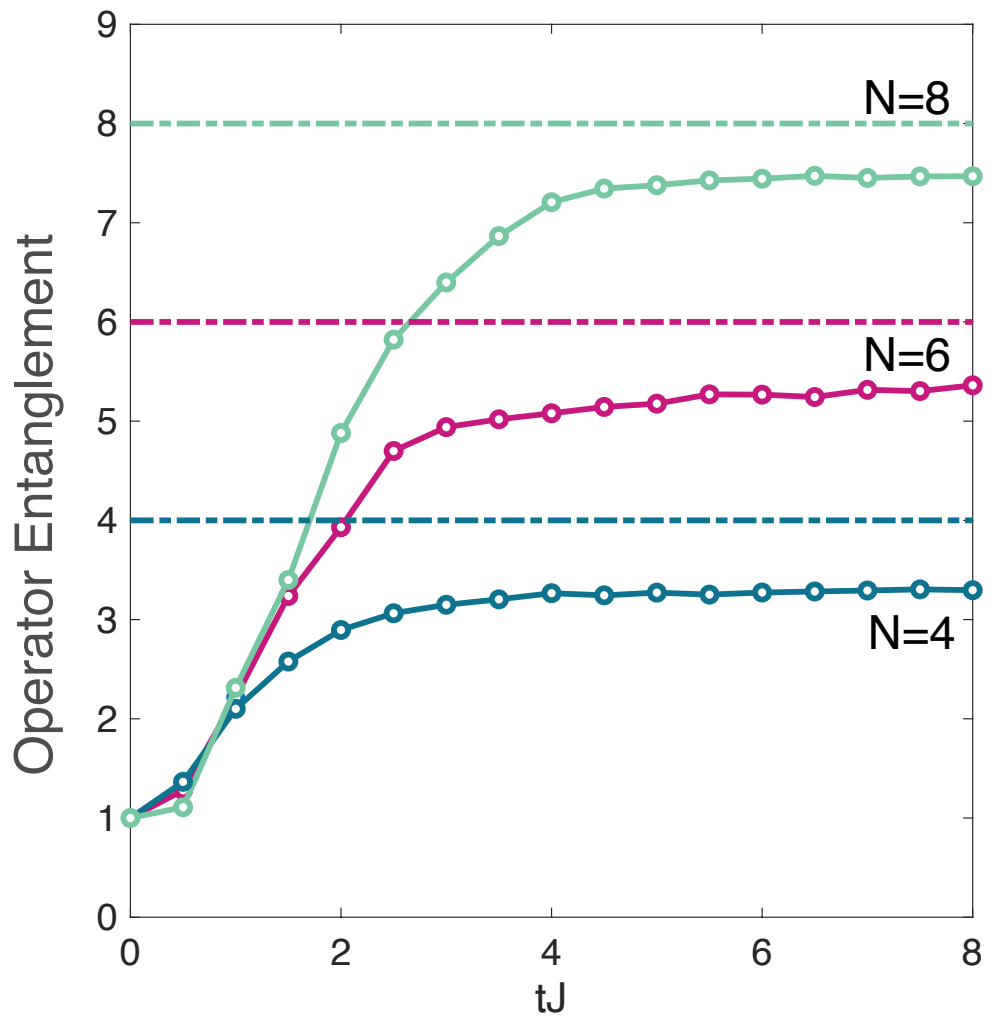

Figure S9: **Operator Entanglement II.** System size scaling of the operator entanglement in the time-evolved $z$-magnetization between a random bi-partition in a random Heisenberg model. The results have been averaged over 100 disorder realizations.

## S5. SOLID-STATE NMR SIMULATION

The effective Hamiltonian for a specific orientation $\Omega$ of a system studied using solid-state NMR can be expressed as

$$H = H_0 + H_\Omega \tag{S.34}$$

where

$$H_0 = \omega_0 S^z_{\text{tot}} \tag{S.35}$$

with represents the global Zeeman term corresponding to the external magnetic field, assumed to be aligned with the z-axis, and

$$H_\Omega = \omega_0 \sum_i \omega_i \mathbf{e}^{(\Omega)}_{\text{cs}} \cdot \mathbf{S}_i + \sum_{i,j} b_{ij} \left(3(\mathbf{S}_i \cdot \mathbf{e}^{(\Omega)}_{\text{ij}})(\mathbf{S}_j \cdot \mathbf{e}^{(\Omega)}_{\text{ij}}) - \mathbf{S}_i \cdot \mathbf{S}_j\right) \tag{S.36}$$

represents the orientation-dependent chemical shifts induced by the external field and direct dipolar interactions between spins [1]. The Zeeman frequency, $\omega_0 = -\gamma_i B_{\text{ext}}$, depends on the gyromagnetic ratios, $\gamma_i$, of the spins and the strength, $B$, of the external magnetic field. For illustration purposes, we assume a homonuclear system where $\gamma \equiv \gamma_i$ is the same for all spins. We allow the external field pointing in the z-direction to define the lab frame and can express the orientation of the system as $\Omega = (\alpha, \beta, \gamma)$, where $\alpha$, $\beta$, and $\gamma$ are Euler angles that define the rotation of the system frame to the lab frame. The chemical shifts, with magnitudes $\omega_0\omega_i$ and directions $\mathbf{e}^{\Omega}_{\text{cs}}$, are most naturally derived starting from the principal axis frame of spin $i$. In this frame, the magnetic field induced in around the spin is parallel to the direction of the external field:

$$\mathbf{B}^{(P)}_i = \overline{\delta}^{(P)}_i \cdot \mathbf{B}^{(P)}_{\text{ext}} \tag{S.37}$$

where

$$\overline{\delta}^{(P)}_i = \begin{bmatrix} \delta^{XX}_i & 0 & 0 \\ 0 & \delta^{YY}_i & 0 \\ 0 & 0 & \delta^{ZZ}_i \end{bmatrix} \tag{S.38}$$

and $\mathbf{B}^{(P)}_{\text{ext}} = B_{\text{ext}} \mathbf{e}^{(P)}_{\text{ext}}$ is the external field expressed in the principal axis system. We can then apply the rotation matrices $\overline{R}(\Omega_i)$ and $\overline{R}(\Omega)$ to move from the principal axis frame to the system frame and from the system frame to the lab frame respectively. We can then write the induced field in the lab frame as

$$\mathbf{B}_i = B_{\text{ext}} \overline{\delta}_i \cdot \mathbf{e}_{\text{ext}} \tag{S.39}$$

where

$$\overline{\delta}^{(\Omega)}_i = \overline{R}^\dagger(\Omega) \cdot \overline{R}^\dagger(\Omega_i) \cdot \overline{\delta}^{(P)}_i \cdot \overline{R}(\Omega_i) \cdot \overline{R}(\Omega) \tag{S.40}$$

$$\mathbf{e}_{\text{ext}} = \overline{R}^\dagger(\Omega) \cdot \overline{R}^\dagger(\Omega_i) \cdot \mathbf{e}^{(P)}_{\text{ext}} \cdot \overline{R}(\Omega_i) \cdot \overline{R}(\Omega). \tag{S.41}$$

Note that $\Omega$ is determined by setting $\mathbf{e}_{\text{ext}} = \mathbf{e}_z$. Recognizing that $\overline{\delta}^{(\Omega)}_i \cdot \mathbf{e}_{\text{ext}}$ is a vector, we define its magnitude and direction as $\omega_i$ and $\mathbf{e}^{(\Omega)}_{\text{cs}}$ respectively via

$$\omega_i \mathbf{e}^{(\Omega)}_{\text{cs}} \equiv \overline{\delta}^{(\Omega)}_i \cdot \mathbf{e}_z. \tag{S.42}$$

The Hamiltonian term resulting from the induced field on spin $i$ is

$$H^{\text{cs}}_i = -\gamma_i \mathbf{B}_i \cdot \mathbf{S}_i = \omega_0 \omega_i \mathbf{e}^{(\Omega)}_{\text{cs}} \cdot \mathbf{S}_i, \tag{S.43}$$

which we see depends on the system orientation $\Omega$. The second term in Eq. (S.34) is a direct dipolar interaction between every pair of spins determined by the dipole vector $\mathbf{e}^{(\Omega)}_{ij}$, which also depends on the system orientation $\Omega$, and the dipolar coupling constant

$$b_{ij} = -\frac{\mu_0 \gamma^2 \hbar}{4\pi r^3_{ij}}, \tag{S.44}$$

which depends on the distance $r_{ij}$ between the spins.

In typical high-field NMR experiments, the energy scale $\omega_0$ of the global Zeeman field is much larger than the other parts of the Hamiltonian. It is common practice, therefore, to perform the secular approximation after moving the frame rotating with $H_0$ to get an effective Hamiltonian

$$\tilde{H} = \omega_0 \sum_i \omega_i \left( \mathbf{e}_{\mathrm{cs}}^{(\Omega)} \cdot \mathbf{e}_z \right) S_i^z + \sum_{i,j} d_{ij} \left( 2\hat{S}_i^z \hat{S}_j^z - \frac{1}{2} \left( \hat{S}_i^+ \hat{S}_j^- + \hat{S}_i^- \hat{S}_j^+ \right) \right) \tag{S.45}$$

The secular dipolar coupling constant is

$$d_{ij} = -\frac{\mu_0 \gamma^2 \hbar}{4\pi r_{ij}^3} \frac{1}{2} \left( 3\cos^2 \Theta_{ij} - 1 \right) \tag{S.46}$$

where $\Theta_{ij}$ is the angle between the dipole vector between the spins and the z-axis. Magic angle spinning of the sample at frequency $\omega_r$ around an axis at a polar angle $\theta_r$ with respect to the z-axis can be included by rewriting the Hamiltonian as [42]

$$\begin{aligned} \tilde{H}_{\mathrm{MAS}} &= \omega_0 \sum_i \left( 1 - \delta_i^{\mathrm{iso}} \right) S_i^z \qquad \text{(S.47)} \\ &+ \omega_0 \sum_{n=-2:2} d_{n,0}^{(2)} \left( \theta_r \right) e^{in\omega_r t} \sum_i \delta_i^{\mathrm{aniso}} \tilde{g}_{n,i} S_i^z \\ &+ \sum_{n=-2:2} d_{n,0}^{(2)} \left( \theta_r \right) e^{in\omega_r t} \sum_{i,j} b_{ij} \tilde{G}_{n,ij} \left( 2\hat{S}_i^z \hat{S}_j^z - \frac{1}{2} \left( \hat{S}_i^+ \hat{S}_j^- + \hat{S}_i^- \hat{S}_j^+ \right) \right) \end{aligned}$$

where we have defined the quantities

$$\delta_i^{\mathrm{iso}} = \frac{1}{3} \left( \delta_i^{XX} + \delta_i^{YY} + \delta_i^{ZZ} \right) \tag{S.48}$$

$$\delta_i^{\mathrm{aniso}} = \delta_i^{ZZ} - \delta_i^{\mathrm{iso}} \tag{S.49}$$

$$\eta_i = \frac{\delta_i^{YY} - \delta_i^{XX}}{\delta_i^{\mathrm{aniso}}} \tag{S.50}$$

$$\tilde{g}_n = \sqrt{\frac{3}{2}} D_{0,n}^{(2)} \left( \alpha_i, \beta_i, \gamma_i \right) - \frac{1}{2} \eta_i \left( D_{2,n}^{(2)} \left( \alpha_i, \beta_i, \gamma_i \right) + D_{-2,n}^{(2)} \left( \alpha_i, \beta_i, \gamma_i \right) \right) \tag{S.51}$$

$$\tilde{G}_{n,ij} = D_{0,n}^{(2)} \left( 0, \beta_{ij}, \gamma_{ij} \right). \tag{S.52}$$

The geometric factors $d_{n,0}^{(2)}$, $D_{0,n}^{(2)}$, and $D_{\pm 2,n}^{(2)}$ depend on rotor angle $\theta_r$, the Euler angles $(\alpha_i, \beta_i, \gamma_i)$ relating the principal axis frame of each spin to the rotor frame, and the Euler angles $(0, \beta_{ij}, \gamma_{ij})$ relating the dipole vector between spins $i$ and $j$ to the rotor frame.

In solid-state NMR experiments, a sample consists of an ensemble of systems, each with a different orientation. The orientations, $\Omega$, takes a uniformly random but static value during the experiment. We can therefore simulate solid-state NMR and experiments by computing the spectrum for an ensemble of Hamiltonians, Eq. (S.45) or Eq. (S.47), each corresponding to a different orientation $\Omega_k$. Performing the ensemble average over orientations, known as the powder average, is onerous for classical computers as we must do several parallel simulations of instances of the dynamics corresponding to fixed orientations, each of which can be time-consuming for large systems. In the main text, we discuss how a powder-averaged simulation performed on a quantum device has roughly the same resource cost for such systems as the simulation of a single instance of the dynamics.